\documentclass[useAMS,usegraphicx,usenatbib]{mn2e} 
\usepackage{longtable}
\usepackage{natbib}
\UseRawInputEncoding
\usepackage{amsmath}
\usepackage{color}
\usepackage{url}
\usepackage{ulem}
\usepackage{multirow}
\usepackage{amsmath}
\usepackage{wasysym}
\usepackage{amssymb}
\usepackage{pifont}

\bibpunct{(}{)}{;}{a}{}{,}

\title[A hard X-ray view of U/LIRGs in GOALS: I]{A hard X-ray view of Luminous and Ultra-luminous Infrared Galaxies in GOALS: I -- AGN obscuration along the merger sequence}
\author[C. Ricci et al.]{C. Ricci$^{1,2}$\thanks{E-mail:
claudio.ricci@mail.udp.cl}, G. C. Privon$^{3}$, R. W. Pfeifle$^{4}$, L. Armus$^{5}$, K. Iwasawa$^{6,7}$, N. Torres-Alb\`a$^{8}$,\newauthor  S. Satyapal$^{4}$, F. E. Bauer$^{9,10,11}$, E. Treister$^{9}$, L. C. Ho$^{2,12}$, S. Aalto$^{13}$, P. Ar\'evalo$^{14}$,\newauthor  L. Barcos-Mu\~noz$^{3}$,V.  Charmandaris$^{15,16}$, T. Diaz-Santos$^{15,16}$, A. S. Evans$^{17,3}$, \newauthor T. Gao$^{18}$, H. Inami$^{19}$,M. J. Koss$^{20}$, G. Lansbury$^{21}$, S. T. Linden$^{22}$, A. Medling$^{23,24}$, \newauthor D. B. Sanders$^{25}$, Y. Song$^{17}$, D. Stern$^{26}$, V. U$^{27}$, Y. Ueda$^{28}$, S. Yamada$^{28}$\\
Affiliations can be found after the references.
 }
\begin{document}
\date{Received; accepted}

\pagerange{\pageref{firstpage}--\pageref{lastpage}} \pubyear{2017}

\maketitle

\label{firstpage}

\begin{abstract}

The merger of two or more galaxies can enhance the inflow of material from galactic scales into the close environments of Active Galactic Nuclei (AGN), obscuring and feeding the supermassive black hole (SMBH). Both recent simulations and observations of AGN in mergers have confirmed that mergers are related to strong nuclear obscuration. However, it is still unclear how AGN obscuration evolves in the last phases of the merger process. We study a sample of 60 Luminous and Ultra-luminous IR galaxies (U/LIRGs) from the GOALS sample observed by {\it NuSTAR}. We find that the fraction of AGN that are Compton-thick (CT; $N_{\rm H}\geq 10^{24}\rm\,cm^{-2}$) peaks at $74_{-19}^{+14}\%$ at a late merger stage, prior to coalescence, when the nuclei have projected separations of $d_{\rm sep}\sim 0.4-6$\,kpc. A similar peak is also observed in the median $N_{\rm H}$ [$(1.6\pm0.5)\times10^{24}\rm\,cm^{-2}$]. The vast majority ($85^{+7}_{-9}\%$) of the AGN in the final merger stages ($d_{\rm sep}\lesssim 10$\,kpc) are heavily obscured ($N_{\rm H}\geq 10^{23}\rm\,cm^{-2}$), and the median $N_{\rm H}$ of the accreting SMBHs in our sample is systematically higher than that of local hard X-ray selected AGN, regardless of the merger stage. This implies that these objects have very obscured nuclear environments, with the $N_{\rm H}\geq 10^{23}\rm\,cm^{-2}$ gas almost completely covering the AGN in late mergers. CT AGN tend to have systematically higher absorption-corrected X-ray luminosities than less obscured sources. This could either be due to an evolutionary effect, with more obscured sources accreting more rapidly because they have more gas available in their surroundings, or to a selection bias. The latter scenario would imply that we are still missing a large fraction of heavily obscured, lower luminosity ($L_{2-10}\lesssim 10^{43}\rm\,erg\,s^{-1}$) AGN in U/LIRGs.

\end{abstract}	
               

   
\section{Introduction}

The discovery of a correlation between the mass of supermassive black holes (SMBHs) and several properties of their host galaxies (e.g., \citealp{Magorrian:1998ly,Ferrarese:2000mx,Gebhardt:2000id,Kormendy:2013fk}) has suggested that the growth of SMBHs and their host galaxies are tightly connected. Mergers of galaxies are thought to be one of the most important mechanisms with which galaxies build up their stellar masses \citep{White:1978nx}. Both observational (e.g., \citealp{Lonsdale:1984sg,Joseph:1985fc,Armus:1987hz,Clements:1996ir,Alonso-Herrero:2000tn,Ellison:2008fv}) and theoretical (e.g., \citealp{Mihos:1996bs}, \citealp{Di-Matteo:2007ly}) studies have shown that galaxy mergers enhance star-formation. Simulations have also shown that the interaction between two or more galaxies can reduce the angular momentum of the circumnuclear material (e.g., \citealp{Barnes:1991mh,Blumenthal:2018ac}), thus providing an effective mechanism to trigger accretion onto SMBHs (e.g., \citealp{Di-Matteo:2005qf}). Observationally, several works have confirmed this scenario. \citet{Koss:2010ve} and \cite{Silverman:2011tg} found a higher AGN fraction in pairs than in isolated galaxies with similar stellar masses. It has been shown that the fraction of AGN in mergers tends to increase as the separation between the two galaxies decreases \citep{Ellison:2011ij}, and peaks after coalescence \citep{Ellison:2013hc}. \citet{Koss:2012qf} have shown that the average luminosity of dual AGN also increases with decreasing separation (see also \citealp{Hou:2020az}), and it is higher for the primary (i.e. more massive) component of the system (see also \citealp{De-Rosa:2019ck} for a recent review). While AGN with moderate X-ray luminosities are typically found in non-interacting disk galaxies (e.g., \citealp{Koss:2011uy,Schawinski:2012yg,Kocevski:2012vn}), more luminous objects are commonly found in merging systems (e.g., \citealp{Treister:2012vn,Hong:2015uq,Glikman:2015lk}). \cite{Treister:2012vn} showed that, while for 2--10\,keV AGN luminosities of $L_{2-10}\sim 10^{41}\rm\,erg\,s^{-1}$ only a small fraction ($< 1\%$) of AGN are in mergers, at $L_{2-10}\sim 10^{46}\rm\,erg\,s^{-1}$ $\sim$70-80\% of the sources are found in interacting systems (see also \citealp{Glikman:2015lk}). Recent evidence has suggested that Hot Dust Obscured Galaxies (Hot DOGs, \citealp{Wu:2012bh,Assef:2015zr}), which are some of the most luminous galaxies observed so far ($L_{\rm\,IR} > 10^{13} L_{\odot}$), are also found in mergers (e.g., \citealp{Fan:2016sf}). These observations suggest that, while at low luminosities SMBH accretion is triggered by secular processes, at high luminosities mergers can play a dominant role. This is in agreement with the evolutionary scenario proposed by \citet{Sanders:1988kl} for ultra-luminous [$L_{\rm\,IR}(8-1000\,\mu\rm m)\geq 10^{12}L_{\odot}$] infrared galaxies (ULIRGs; e.g., \citealp{Sanders:1996uq}, \citealp{Perez-Torres:2021wa}). In this scheme two gas-rich disk galaxies collide, triggering star-formation and accretion onto the SMBH. The strong accretion onto the SMBH would lead the source to evolve first in a luminous red quasar (e.g., \citealp{Urrutia:2008vn,Glikman:2015lk,LaMassa:2016ly}) and then in an unobscured blue quasar.

The bulk of the growth of SMBHs during mergers is believed to be very obscured. This has been shown by numerical simulations (e.g., \citealp{Hopkins:2008xr,Blecha:2018gt,Kawaguchi:2020qb}), as well as by recent observations. \cite{Satyapal:2014oq} have shown that post-mergers host a significantly higher fraction of mid-IR selected AGN than optical AGN, which could suggest that optically obscured AGN become prevalent in the most advanced mergers (see also \citealp{Koss:2010ve,Ellison:2019ve,Secrest:2020ik}). \cite{Kocevski:2015zr} have shown that heavily obscured ($N_{\rm\,H}\geq 10^{23.5}\rm\,cm^{-2}$) systems are more common in mergers than in isolated galaxies.  In the local Universe major galaxy mergers give rise to Luminous Infrared Galaxies [LIRGS; $L_{\rm\,IR}(8-1000\,\mu\rm m)=10^{11}-10^{12}$ $L_{\odot}$] and ULIRGs which, over the past two decades, have been extensively studied in the IR, optical and soft (0.3--10\,keV) X-ray bands (e.g., \citealp{Veilleux:1995if,Veilleux:1999kc}; \citealp{Imanishi:2000fu,Imanishi:2002qf,Imanishi:2006on}; \citealp{Alonso-Herrero:2006hw,Alonso-Herrero:2012ud,Armus:2007gb,Armus:2009fk,Armus:2020kh,Teng:2010fk}; \citealp{Franceschini:2003uq,Pereira-Santaella:2011fu,Nardini:2011tn}).  Mid-IR observations have suggested the presence of a heavily buried AGN in U/LIRGs, particularly in those undergoing the final stages of mergers (e.g., \citealp{Imanishi:2007hl,Veilleux:2009qo,Nardini:2010wl}). Hard X-ray ($\geq 10$\,keV) observations can be extremely effective in detecting heavily obscured AGN and, combined with soft X-ray observations ($<10$\,keV), in estimating their line-of-sight column density (e.g., \citealp{Burlon:2011wl,Ricci:2015fk,Ricci:2017if,Annuar:2015qf,Koss:2016pk,Koss:2016oz}). U/LIRGs were studied in the hard X-ray band using {\it Swift}/BAT by \citet{Koss:2013pi}, who suggested that a large fraction of sources might have CT column densities. Exploiting the revolutionary capabilities of {\it NuSTAR}, the first focussing hard X-ray satellite on orbit, \citet{Ricci:2017aa} studied 30 nearby U/LIRGs from the Great Observatories All-sky LIRG Survey (GOALS, \citealp{Armus:2009fk}) sample\footnote{http://goals.ipac.caltech.edu/}. \citet{Ricci:2017aa} showed that $65^{+12}_{-13}\%$ of the AGN in objects in late-stage mergers (i.e. with projected separations of $d_{\rm sep}\simeq 10$\,kpc) are Compton-thick (CT, $N_{\rm H}\geq 10^{24}\rm\,cm$), a fraction significantly higher than what is found for local hard X-ray selected AGN ($27\pm4\%$, \citealp{Ricci:2015fk}), which are typically found in non-interacting systems. Similar results have also been found by several other studies, which find that AGN in mergers are systematically more obscured than those in isolated galaxies (e.g., \citealp{Nardini:2011tn,Lanzuisi:2015oy,Del-Moro:2016va,Koss:2016pk,Koss:2018iv,Satyapal:2017mk,Dutta:2018zn,Dutta:2019bn,Goulding:2018nc,Donley:2018sr,Pfeifle:2019vd,Pfeifle:2019gb,Secrest:2020ik,Foord:2021ox,Guainazzi:2021nc}). At higher luminosities and redshifts, X-ray observations of Hot DOGs have shown that these powerful AGN are also typically very obscured (e.g., \citealp{Piconcelli:2015jn,Ricci:2017pd,Zappacosta:2018vn,Toba:2020tf}).

\begin{figure}
\centering
\includegraphics[width=8.5cm]{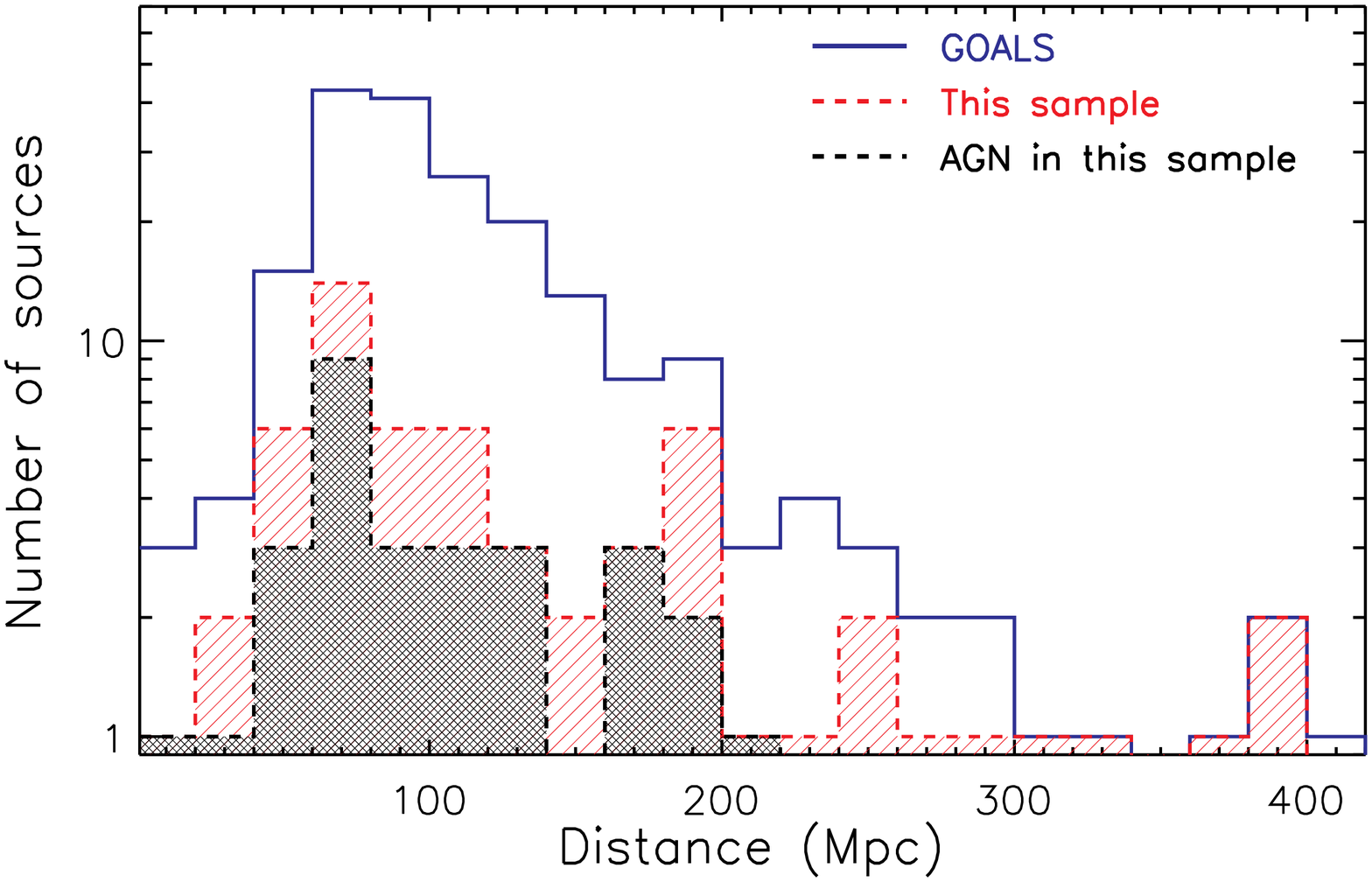}
\smallskip
\includegraphics[width=8.5cm]{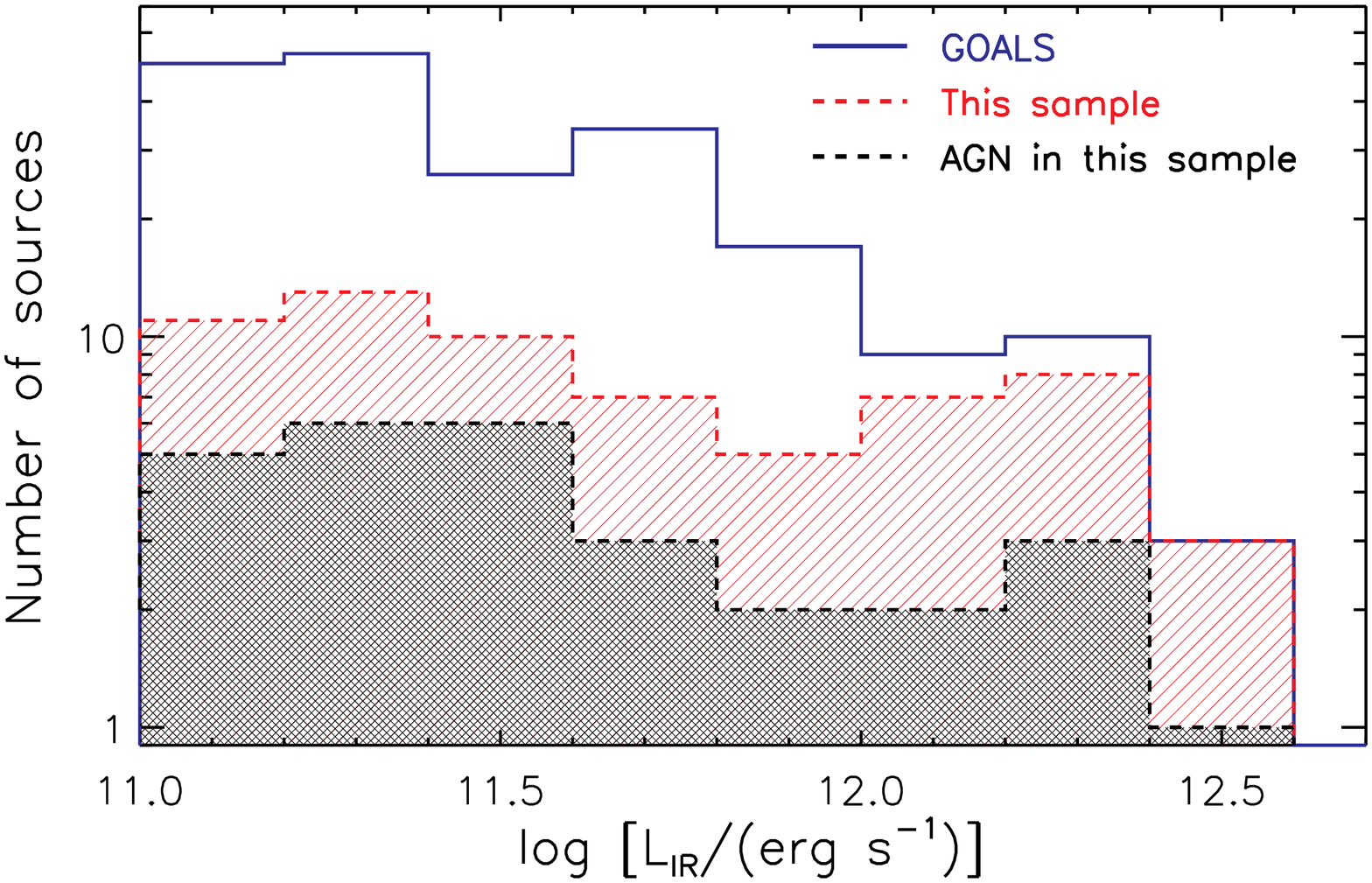}
\smallskip
\includegraphics[width=8.5cm]{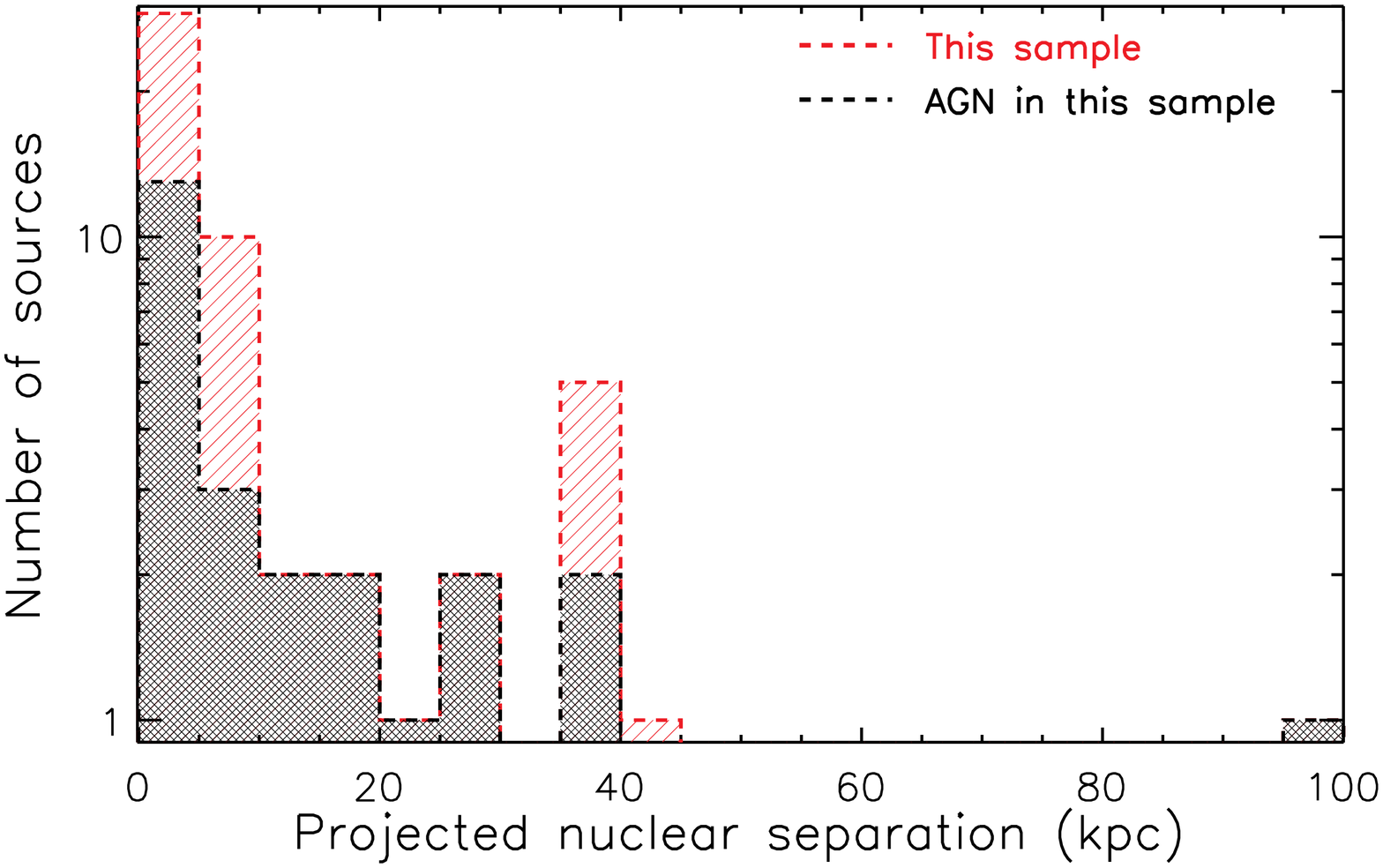}
  \caption{{\it Top panel:} Histogram of the distance to the objects in the GOALS sample (blue continuous line), to our sample (red dashed line), and to the X-ray AGN in our sample (black dashed line). {\it Middle panel:} Histogram of the $8-1000\,\mu$m IR luminosity. {\it Bottom panel:} Histogram of the projected separation between the two nuclei of the merging galaxies. }
\label{fig:histograms}
\end{figure}

While a growing number of observations have demonstrated that the obscuration properties of AGN in mergers are very different from those of AGN in isolated galaxies (see \citealp{Ramos-Almeida:2017lk,Hickox:2018ur} for recent reviews), it is still unclear how AGN obscuration evolves in the last phases of the merger process, when the two nuclei are at a projected separation of $d_{\rm sep}< 10$\,kpc. With the goal of addressing this important issue, and to increase the number of sources with $d_{\rm sep}< 10$\,kpc, in this work we double, with respect to \citet{Ricci:2017aa} the number of U/LIRGs from the GOALS sample observed in the hard X-rays by {\it NuSTAR}. 
GOALS is a sample of nearby ($z<0.088$) galaxies detected by the {\it Infrared Astronomical Satellite} ({\it IRAS}) revised bright Galaxy Survey \citep{Sanders:2003fk}, which has a very wealthy collection of ancillary data across the whole multi-wavelength spectrum (e.g., \citealp{Howell:2010uq,Petric:2011zt,Stierwalt:2013ff}). Exploiting the excellent constraints on the AGN obscuration obtained by broad-band X-ray observations, we study here the relation between obscuration and merger stage, focussing in particular on the final stages of the merger process. A companion paper \citep{Yamada:2021xu} focuses on the physical X-ray modelling of these sources, to constrain the covering factor of the torus from X-ray spectroscopy, and on the X-ray to bolometric AGN luminosity ratios, to further discuss their nuclear properties in comparison with normal AGNs. 

The paper is structured as follows. In $\S$\ref{sec:Sample} we describe our sample, in $\S$\ref{sec:datareduction} we present the X-ray data used and the methodology for the data reduction.  In $\S$\ref{sec:xrayspec} we discuss the spectral analysis of the sources. In \S\ref{sec:agnobscuration} we discuss the relation between mergers and AGN obscuration. Our main results are summarized in \S\ref{sec:summary}. 
Throughout the paper we adopt standard cosmological parameters ($H_{0}=70\rm\,km\,s^{-1}\,Mpc^{-1}$, $\Omega_{\mathrm{m}}=0.3$, $\Omega_{\Lambda}=0.7$). Unless otherwise stated, uncertainties are quoted at the 90\% confidence level.

\begin{table*}
\centering
\caption{Sample of 60 U/LIRGs from GOALS with {\it NuSTAR} observations. (1) IRAS name, (2) counterparts, (3) redshift, (4) merger stage, (5) projected separation between the two nuclei in arcsec and (6) in kpc, (7) star-formation rate estimated from the IR luminosity excluding AGN contribution, and (8)  $8-1000\,\mu$m IR luminosity. In (5) and (6) we report ``S" for objects for which a single nucleus is observed. The sources classified as $a$ and $b$ are early-stage mergers, while those in $c$ and $d$ are late-stage mergers. Sources in the $N$ class are those who do not show any clear sign of merger. In the objects in which more than one SFR or IR luminosity are reported we listed the values for both nuclei.}\label{tab:sample} 
\begin{center}
\begin{tabular}{llcccccc}
\noalign{\smallskip}
\hline \hline \noalign{\smallskip}
\multicolumn{1}{c}{(1)}  & \multicolumn{1}{c}{(2)} & (3) & (4) & (5) & (6)  & (7) & (8)  \\
\noalign{\smallskip}
{\it IRAS} name & \multicolumn{1}{c}{Source} & z  & M &  $d_{\rm sep}$  & $d_{\rm sep}$ & SFR  & $\log (L_{\rm IR}/L_{\odot})$ \\
\noalign{\smallskip}
 & \multicolumn{1}{c}{ } &    &   &  $\arcsec$ &  kpc & $\rm M_{\odot}\,\rm yr^{-1}$ & \\
\noalign{\smallskip}
\hline \noalign{\smallskip}
\noalign{\smallskip}
F00085$-$1223  &   NGC\,34                               &  0.0196   & $d$     &  S     & S  &      44.2  &		11.49	   \\      
\noalign{\smallskip}	
F00163$-$1039  &   Arp\,256  (MCG$-$02$-$01$-$051   \& MCG$-$02$-$01$-$052)                   & 0.0272       & $b$    &  64.3     &  37.1  &    37.8/4.2  &		11.44/10.45	   \\    
\noalign{\smallskip}	
F00344$-$3349  & ESO\,350$-$IG038   &  0.0206  & $c$    &   3.1    &    1.1  &      21.8  &		11.28	   \\
\noalign{\smallskip}	
 F00506+7248 &   MCG+12$-$02$-$001           &  0.0157  & $c$    & 0.9      & 0.3  &    43.7    &		11.50		  \\         
\noalign{\smallskip}	
F01053$-$1746 & VV\,114 (IC\,1623A \& IC\,1623B) &  0.0203  & $c$    &   10.5    &     4.5  &      54.6   &		11.62	 \\
\noalign{\smallskip}	
F02069$-$1022  &   NGC\,833	\&  NGC\,835			& 0.0129  & $a$    &   55.9   & 15.7      &    7.8/1.7  &	10.80/10.02 		  \\  
\noalign{\smallskip}	
F02401$-$0013  &   NGC\,1068&  0.0038  &   $N$  &    --   &  --     &       17.3     &		11.40	     \\
\noalign{\smallskip}	
F03117+4151   &  Mrk\,1073	& 0.0233   & $N$   & --      & --     &       19.2 &	11.41		   \\
\noalign{\smallskip}	
F03164+4119  &   NGC\,1275            &   0.0176  &  $N$  & --      & --    &    15.1  &	11.26		      \\
\noalign{\smallskip}	
F03316$-$3618 &    NGC\,1365	&  0.0055  & $N$  &  --     &     --  &     17.9  &	11.00		   \\
\noalign{\smallskip}	
F04454$-$4838 &ESO\,203$-$IG001   &   0.0529 & $b$    &    7.4   &    8.5  &       68.3  &	11.86		  \\
\noalign{\smallskip}	
F05054+1718  &   CGCG 468$-$002 (E \& W)				&  0.0182  & $b$     &  29.7    & 11.3 &   15.4/4.0   &		11.03/10.72	      \\
\noalign{\smallskip}	
F05189$-$2524 &  IRAS\,05189$-$2524			& 0.0426   & $d$    &    S    & S   &        86.1	 &		12.16	  \\  
\noalign{\smallskip}	
F06076$-$2139    & IRAS\,F06076$-$2139 (S \& N) & 0.0375   & $c$   &   7.8    &     6.2 &    51.8/7.1   &		11.59/10.73	    \\
\noalign{\smallskip}	
07251$-$0248	&  &  0.0875   & $d$   &    S    &   S    &   315    &	12.39		    \\
\noalign{\smallskip}	
F08354+2555 & NGC\,2623   & 0.0185  & $d$    &  S     &    S   &   52.8   &		11.59	    \\
\noalign{\smallskip}	
F08520$-$6850  &   ESO\,060$-$IG16    (NE \& SW)       & 0.0463   & $c$  &  9.4     &    9.4   &      64.0/11.6  &	11.75/11.00		   \\  
\noalign{\smallskip}	
F08572+3915  &   IRAS\,08572+3915 (NW \& SE)						&  0.0584  &  $d$    &    4.4    & 5.6  &     114.9   &	12.16			  \\   
\noalign{\smallskip}	
F09111$-$1007	&    	IRAS\,F09111$-$1007	(W \& E)		&   	0.0541  &  $b$    &  36.4    & 43.4  &  136/38      &	11.96/11.40		  \\ 
\noalign{\smallskip}	
F09320+6134  &   UGC\,05101							&  0.0394  & $d$     &   S     &  S &       114.7  &		12.01		  \\  
\noalign{\smallskip}	
F09333+4841  &   MCG+08$-$18$-$013   \& MCG+08$-$18$-$012 	&  0.0259 & $a$ &   65.6    &  36.0 &     24.7/1.4  &	11.32/9.98		   \\    
 \noalign{\smallskip}	
 F10015$-$0614 &   NGC\,3110     \& MCG$-$01$-$26$-$013              	& 0.0169	 & $a$   &  108.9    &  37.7  &     31.9/3.9 &	11.38/10.42		      \\             
\noalign{\smallskip}	
F10038$-$3338   &  IRAS\,F10038$-$3338       &  0.0341  & $d$  & S      &     S  &        70.3  &		11.78	\\
\noalign{\smallskip}	
 F10257$-$4339  &   NGC\,3256					& 0.0094   & $d$    &   5.1   & 1.0   &  61.1     &		11.64	 	 \\  
\noalign{\smallskip}	
F10565+2448  &   IRAS\,10565+2448W						& 0.0431   &  $c$      &   7.4   &  6.7  &     172.5   &	12.08			  \\  
\noalign{\smallskip}	
F11257+5850 &   Arp\,299	(NGC3690W \& NGC3690E)							& 0.0102   & $c$    &    22.2  &   4.7&      50.4/53.9  &		11.67/11.58		 \\
\noalign{\smallskip}	
 F12043$-$3140 &   ESO\,440$-$IG058 (N \& S)                &  0.0234      & $b$   &   13.4    &  6.6  &  4.6/33.6  &	10.49/11.38		     \\          
\noalign{\smallskip}	
F12112+0305   & &  0.0733  & $c$   &    3.5   &  5.6 &      322.5  &		12.36	   \\
\noalign{\smallskip}	
F12243$-$0036  &   NGC\,4418              &  0.0073  &   $N$&   --    &  --   &     11.9  &		11.19	     \\
\noalign{\smallskip}	
  F12540+5708 &   Mrk\,231									&0.0422    & $d$     &   S    & S  &     259.7  &	12.57		     \\   
\noalign{\smallskip}	
F12590+2934 &   NGC\,4922 (N \& S)                          &  0.0232  & $c$    &  22.3    &     10.9 &     29.2/0.48   &		11.37/9.51		 \\      
\noalign{\smallskip}	
 13120$-$5453  &   IRAS\,13120$-$5453						&  0.0308  & $d$     &  S     & S   &   299.4   &	12.32		    \\  
\noalign{\smallskip}	
F13126+2453  &   IC\,860                 &  0.0112  & $N$  &   --    &  --    &    19.1  &	11.14		     \\
\noalign{\smallskip}	
F13188+0036      & NGC\,5104 &  0.0186  & $N$   &   --    &     --  & 24.7      &11.27   \\
\noalign{\smallskip}	
F13197$-$1627	&	MCG$-$03$-$34$-$064 \& MCG$-$03$-$34$-$063										&	0.0213		& $a$	&   106.4   &  37.8&     2.7/6.2 &	11.17/10.61		 	\\
\noalign{\smallskip}	
F13229$-$2934   &  NGC\,5135	&  0.0137   & $N$  &   --    &  --   &      22.5    &	11.30		  \\
\noalign{\smallskip}	
F13362+4831 &	NGC\,5256 (SW \& NE) & 0.0279   & $c$  &  10.2     &    6.0   &    25.8/15.4  &	11.35/11.13		    \\
\noalign{\smallskip}	
F13428+5608  &   Mrk\,273								& 0.0378    &  $d$    &    0.9  & 0.7    &    166.0    &		12.21	   \\   
\noalign{\smallskip}	
F14348$-$1447   & F14348$-$1447 (NE \& SW)  & 0.0830   & $c$    &    4.0   &     7.3 &     327.1  &		12.383	     \\
\noalign{\smallskip}	
F14378$-$3651  &   IRAS\,14378$-$3651						&  0.0676  & $d$       &    S   &  S  &    238.5   &		12.23	     \\  
\noalign{\smallskip}	
F14544$-$4255  &   IC\,4518A	\& IC\,4518B					& 0.0163   & $b$    &   44.7   &   15.3  &    21.5/4.0  &	11.16/10.43		    \\  
\noalign{\smallskip}	
F15250+3608  & &  0.0552  & $d$    &   S    &    S &   146.1   &		12.08	      \\
\noalign{\smallskip}
\hline
\noalign{\smallskip}
\end{tabular}
\end{center}
\end{table*} 

\setcounter{table}{0}
\begin{table*}
	\centering
	\caption{Continued. }
\begin{center}
\begin{tabular}{llcccccc}
\noalign{\smallskip}
\hline \hline \noalign{\smallskip}
\multicolumn{1}{c}{(1)}  & \multicolumn{1}{c}{(2)} & (3) & (4) & (5) & (6)  & (7) & (8)  \\
\noalign{\smallskip}
{\it IRAS} name & \multicolumn{1}{c}{Source} & z  & M &  $d_{\rm sep}$  & $d_{\rm sep}$ & SFR  & $\log (L_{\rm IR}/L_{\odot})$ \\
\noalign{\smallskip}
 & \multicolumn{1}{c}{ } &    &   &  $\arcsec$ &  kpc & $\rm M_{\odot}\,\rm yr^{-1}$ & \\
\noalign{\smallskip}
\hline \noalign{\smallskip}
\noalign{\smallskip}	
F15327+2340  &   Arp\,220 (W \& E)								& 0.0181   & $d$     &   1.0   & 0.4   &       254.1 &		12.27		  \\   
\noalign{\smallskip}	
F16504+0228  &   NGC\,6240 (N \& S)											& 0.0245   &  $d$     &   1.4   & 0.7  &     112.1   &		11.93	   	  \\	
\noalign{\smallskip}	
F16577+5900 &   NGC\,6286    \& NGC\,6285                          &  0.0183     &  $b$   &    91.1   & 35.8  	&    26.2/9.8  &		11.30/10.85	     \\    
\noalign{\smallskip}	
F17138$-$1017  &   IRAS\,F17138$-$1017          &  0.0173   & $d$      &    S   & S   &     42.5   &		11.49	    \\     
\noalign{\smallskip}	
F17207$-$0014	 &        &   0.0428  &  $d$  &   S    &   S   &    405.0   &		12.46	    \\ 
\noalign{\smallskip}	
F18293$-$3413  &   IRAS\,F18293-3413     &   0.0182 & $N$  &  --     &    --  &     106.5   &		11.88	   \\ 
\noalign{\smallskip}	
F19297$-$0406  &  &   0.0857 & $d$    &  S     &   S  &   402.1  &		12.45	     \\
\noalign{\smallskip}	
F20221$-$2458      & NGC\,6907 &  0.0106  & $N$   &   --    &  --   &     17.6  &	11.11		     \\
\noalign{\smallskip}	
20264+2533  &   MCG $+$04$-$48$-$002 \& NGC\,6921					& 0.0139   & $a$   &   91.4   & 27.1  &    12.6/7.9  &		11.01/10.73	       \\  
\noalign{\smallskip}	
F20550+1655  & CGCG\,448$-$020 (E \& W)  &   0.0359 & $c$    &  5.0     &     3.8 &    83.0/30.6  &		11.77/11.34	   \\
\noalign{\smallskip}	
F20551$-$4250  &   ESO\,286$-$IG19           & 0.0430    & $d$  &   S    &    S   &    130.9  &		12.06	    \\
\noalign{\smallskip}	
F21453$-$3511  &   NGC\,7130								&0.0162   & $d$     &   S    & S  &    30.3  &		11.42	    \\
\noalign{\smallskip}	
F23007+0836  &   Arp\,298 (NGC\,7469	\& IC\,5283)				& 0.0163   &  $a$    & 79.7     & 26.8  &     43.3/9.2   &		11.58/10.79	 	  \\
\noalign{\smallskip}	
F23128$-$5919  &ESO\,148$-$IG002   &  0.0446  & $c$    &   4.7    &      4.5 &   139.5   &		12.06	   \\
\noalign{\smallskip}	
F23157+0618     & NGC\,7591 &  0.0165  & $N$   &    --   &  --  &   17.7    &		11.11	      \\
\noalign{\smallskip}	
F23254+0830  &   Arp\,182 (NGC\,7674	\& NGC\,7674A	)							& 0.0289  &  $b$   &  34.1    & 20.7  	&  13.5/2.0    &		11.54/10.14	    \\
\noalign{\smallskip}	
 23262+0314 &   NGC\,7679	\& NGC\,7682				& 0.0171  & $a$    &   269.7   & 97.3   &     16.0/--    &		11.11/--	   \\
\noalign{\smallskip}	
F23365+3604	&   			& 0.0645    &  $d$   &  S     &  S     &   224.3  &		12.20	     \\ 
\noalign{\smallskip}
\hline
\noalign{\smallskip}
\end{tabular}
\end{center}
\end{table*}

\section{Sample}\label{sec:Sample}

The all-sky GOALS sample consists of 180 LIRGs and 22 ULIRGs, and is complete at  60$\mu$m for fluxes $>5.24$\,Jy. Objects in GOALS have been extensively studied in the IR, with a large number of observations carried out by {\it Spitzer}, {\it Akari} and {\it Herschel} (e.g., \citealp{Inami:2010pf,Petric:2011zt,Diaz-Santos:2011tg,U:2012fc,U:2019hq,Inami:2013il,Stierwalt:2013ff,Stierwalt:2014dg,Medling:2014ps,Lu:2017mp,Inami:2018mp}). Moreover, a large {\it Chandra} campaign provides spectroscopic coverage in the 0.3--10\,keV range, as well as high spatial resolution images in the same band \citep{Iwasawa:2011fk,Torres-Alba:2018yj}.
Our sample includes all U/LIRGs from the GOALS sample that were observed by {\it NuSTAR}. This includes the 30 objects reported in \cite{Ricci:2017aa}, besides sources that were recently analyzed in literature studies, as well as 19 objects that have been recently observed by {\it NuSTAR} as a part of several observational campaigns led by our team (PIs: Ricci, C; Privon, G.; Armus, L.) to study SMBH accretion in the final phases of the merger process. Overall our sample contains 60 U/LIRGs. 

\subsection{Merger stages}
Near-infared (NIR) and mid-infrared (MIR) images were used to classify the sources into different merger stages. We followed what was reported by \cite{Haan:2011fk} using {\it HST} images and, when that was not available, we considered the classification of \citet{Stierwalt:2013ff}, including the modifications proposed by \citet{Ricci:2017aa}. Based on the morphological properties of the objects, we divided them into five different merger stages, following \citet{Stierwalt:2013ff}: \smallskip\newline
\textbf{Stage \textit{a}}: galaxy pairs before a first encounter.\newline
\textbf{Stage \textit{b}}: galaxies after a first-encounter, with symmetric galaxy disks but showing signs of tidal tails.\newline
\textbf{Stage \textit{c}}: systems showing strong tidal tails, amorphous disks, and other signs of merger activity.\newline
\textbf{Stage \textit{d}}: galaxies in the final merger stages, with the two nuclei being in a common envelope or showing only a single nucleus.\newline
\textbf{Stage \textit{N}}: sources which do not appear to be in a major merger. These sources could either be post-mergers or minor mergers.\smallskip\newline
Sources in the early merger stages are classified as belonging to the $a$ and $b$ class, while those in late stage mergers have been classified as being in the $c$ or $d$ stage (see \citealp{Stierwalt:2013ff} for details, and Fig.\,1 of \citealp{Ricci:2017aa}). Typically sources in late stage mergers are separated by $d_{\rm sep}\lesssim 11$\,kpc. All the sources in our sample, together with their merger stages and the projected distances between the two nuclei, are listed in Table\,\ref{tab:sample}, while in Fig.\,\ref{fig:histograms} we illustrate some of their main properties. The closest observed projected distance for systems showing at least an AGN is $d_{\rm sep} = 0.4$\,kpc, therefore we assign this distance as the minimum distance between two potential AGN in this study. Of the 60 sources in our sample, seven are in stage $a$, eight in stage $b$, 13 in stage $c$, 21 in stage $d$ and 11 in stage $N$. This doubles the number of U/LIRGS with {\it NuSTAR} observations with respect to the sample presented in \cite{Ricci:2017aa}, and in particular we have now observations of 34 late-stage galaxies, while only 17 were reported in \cite{Ricci:2017aa}. 

\subsection{Star formation rates}
The star-formation rates (SFRs) and IR luminosities were taken from \citet{Diaz-Santos:2017va}. The SFRs were obtained based on the host galaxy IR luminosity (excluding the AGN contribution estimated by \citealp{Diaz-Santos:2017va}), using the relation reported by \citet{Murphy:2011tl}. We privileged these values rather than the more recent compilation of \citet{Shangguan:2019eb}, since it allowed us to recover the SFRs for the individual galactic nuclei. We tested the SFRs of \citet{Shangguan:2019eb}, and found results consistent with those we obtained using the aforementioned approach. For three objects in our sample, which were not reported in \citet{Diaz-Santos:2017va}, we used values from recent literature. For NGC\,1068 and NGC\,1365 we used the SFRs obtained by \citet{Ichikawa:2017ec,Ichikawa:2019kg}, while for the Hickson compact Group 16 (HCG16, \citealp{Hickson:1982fm}) we used the values reported in \cite{OSullivan:2014vj} and \cite{Bitsakis:2014bb}.

\subsection{Comparison sample}\label{sec:comparisonBAT}

As a comparison sample, similarly to what was done in \cite{Ricci:2017aa}, we use AGN reported in the {\it Swift}/BAT 70-month catalogue \citep{Baumgartner:2013uq}, which were selected in the 14--195\,keV band. Studying optical images, \cite{Koss:2010ve} showed that only $\sim 25\%$ of the AGN detected by BAT are found in major mergers with a nuclear separation $d_{\rm sep}\lesssim 100$\,kpc. The broad band (0.3--150\,keV) X-ray spectra of these $\sim 840$ AGN have been analysed in detail by \cite{Ricci:2017if}, who reported values of the column density for $\sim 99.8\%$ of them. The obscuration properties of the $\sim 730$ non-blazar AGN in the sample were discussed in \citet{Ricci:2015fk,Ricci:2017ej}, who found that $27\pm4\%$ of the objects are CT, and $70\%$ of them are obscured [$\log (N_{\rm H}/\rm cm^{-2})\geq 22$].

\section{Data reduction}\label{sec:datareduction}
In this work we analyze X-ray data obtained from the {\it NuSTAR}, {\it Chandra} and {\it XMM-Newton} facilities, the data reduction of which we outline in \S\ref{sec:nustardatared}, \S\ref{sec:chandradatared} and \S\ref{sec:XMMdatared}, respectively. The extraction regions of the different instruments were selected to cover the host galaxies. We combine these with similar X-ray data previously analyzed and presented in \citet{Ricci:2017aa} for 30 GOALS U/LIRGs, and literature constraints on several additional objects. The details of all X-ray observations analyzed here are listed in Table\,\ref{tab:obslog} in Appendix\,\ref{sect:appendix_obslog}.

\subsection{NuSTAR}\label{sec:nustardatared}

We analyze {\it NuSTAR} \citep{Harrison:2013uq} observations for 23 sources using the {\it NuSTAR} Data Analysis Software \textsc{nustardas}\,v1.9.2 within \textsc{heasoft\,v6.27}. We adopted the calibration files released on May\,6 2020 \citep{Madsen:2015uq}. In order to extract the source spectra we use a circle of $50\arcsec$, while for the background we consider an annulus centred on the source, with an inner and outer radius of $60\arcsec$ and $100\arcsec$, respectively. In several cases, no X-ray source is detected by {\it NuSTAR}, and for these sources we follow the same approach reported in \cite{Lansbury:2017or} to calculate the flux upper limits. This is done using the Bayesian approach of \citet{Kraft:1991kq}.

\subsection{Chandra}\label{sec:chandradatared}

{\it Chandra}/ACIS \citep{Weisskopf:2000vn,Garmire:2003kx} observations are available for all of the new sources of our sample. We reduce the  observations following standard procedures, using \textsc{CIAO} v.4.10. We reprocess all data sets using the \textsc{chandra\_repro} task, and then extract the spectra using a circular region with a radius of $10\arcsec$. For the background spectra we used a circular region with the same radius, selected in region devoid of other X-ray sources. In the case of IC\,1623B, due to its extended emission, we used a radius of $20\arcsec$, for the source, in order to consider all the X-ray emission from the source, consistent with was done to obtain the {\it NuSTAR} and {\it XMM-Newton} spectra; considering a smaller radius ($10\arcsec$) we obtained similar results for this source (i.e. no clear sign of AGN activity). For IRAS\,14348$-$1447 and IRAS\,20550+1655 we also extracted the X-ray emission from the individual nuclei, considering source regions of $1.8\arcsec$ and $2.0\arcsec$, respectively. Among the new sources of our sample, only ESO\,203$-$IG001 was not detected by {\it Chandra}.

\subsection{XMM-Newton}\label{sec:XMMdatared}

We include {\it XMM-Newton} \citep{Jansen:2001vn} observations for nine sources. The EPIC/PN \citep{Struder:2001uq} spectra are obtained by first reducing the original data files using {\it XMM-Newton} Standard Analysis Software (SAS) version 18.0.0 \citep{Gabriel:2004fk}, and then using the \texttt{epchain} task. We filter all observations to remove periods of high-background activity, by analysing the EPIC/PN background light curve in the 10-12\,keV band. Finally, the spectra is extracted by using a circular region of $25\arcsec$ radius, while the background is extracted on the same CCD, in a region devoid of X-ray sources, using a circular region of $40\arcsec$ radius. None of the observations is significantly affected by pileup.

\section{X-ray spectral analysis}\label{sec:xrayspec}

\subsection{Spectral modelling}\label{sec:xrayspectralmodelling}
We fit the X-ray spectra of all sources starting with a star-formation (SF) model, which consists of a power-law component (\textsc{zpow} in \textsc{xspec}) and a collissionally-ionized plasma (\textsc{apec}). We include Galactic absorption using the \textsc{tbabs} model \citep{Wilms:2000vn}, fixing the column density to the value reported by \cite{Kalberla:2005fk} at the coordinates of the source. Intrinsic absorption is considered by including a \textsc{ztbabs} component.
Overall the star-formation model used is: \textsc{tbabs$\times$ztbabs$\times$(zpow+apec)}. In a few cases, for which the signal-to-noise ratio is particularly low, we use a simple power-law model to fit the spectra [\textsc{tbabs$\times$ztbabs$\times$(zpow)}]. 

\begin{figure}
\centering
\includegraphics[width=8.8cm]{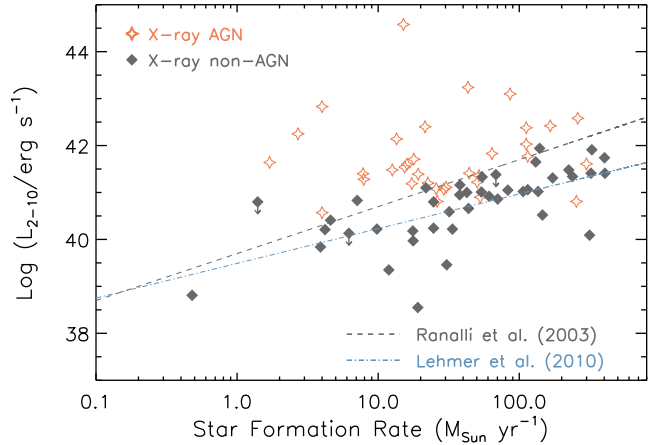}
  \caption{Observed 2--10\,keV X-ray luminosities (see Table\,\ref{tab:Xrayprop}) versus star-formation rate (see Table\,\ref{tab:sample}) for the sources of our sample divided into X-ray AGN (red empty stars) and X-ray non-AGN (black filled diamonds). The X-ray non-AGN have 2--10\,keV luminosities consistent, or lower, than the values expected from their SFR (black continuous line, \citealp{Ranalli:2003yb}; blue dot-dashed line, \citealp{Lehmer:2010hj}), while X-ray AGN are typically more luminous.}
\label{fig:lxvssfr}
\end{figure}

The X-ray spectra of most (21/23) of the sources analyzed here could be well reproduced by a star-formation model. For the two sources (NGC\,2623 and ESO\,060$-$IG16) that show a strong Fe K$\alpha$ line at 6.4 keV, or a clear excess over the star-formation model, we include an AGN component to account for the excess. This is done using the \textsc{RXTorus} model, developed using the \textsc{RefleX} ray-tracing platform \citep{Paltani:2017wq}. The model assumes a toroidal absorber surrounding the accreting system. The inner-to-outer radius ratio of the torus is fixed to $0.5$, while the inclination angle to $90^{\circ}$ (i.e. corresponding to an edge-on scenario). In \textsc{xspec} the model is: \textsc{tbabs$\times$ztbabs$\times$(zpow+apec+atable\{RXTorus\_rprc\_200\}} \textsc{+etable\{RXTorus\_cont\}*zcutoffpl)}, where \textsc{RXTorus\_rprc\_200} and \textsc{RXTorus\_cont} are the reprocessed radiation and obscuration components, while \textsc{zcutoffpl} is a cutoff power-law model used for the continuum. In the latter component the cutoff energy is fixed to 200\,keV \citep{Ricci:2018wb}. The \textsc{zpow} component includes contributions from both star-forming regions and from scattered X-ray emission (e.g., \citealp{Ueda:2007fk,Ueda:2015fh,Ricci:2017if,Gupta:2021vl}). The parameters obtained by our spectral analysis are reported in Table\,\ref{tab:Xray_results1} in Appendix\,\ref{sect:appendix_spectralanalysis}, while the column densities and intrinsic AGN luminosities, for all the objects in our sample, are reported in Table\,\ref{tab:Xrayprop}.

Details on the X-ray spectral analysis of the individual sources are reported in Appendix\,\ref{sect:individualsources}. We verified whether the observed 2--10\,keV luminosity is consistent with what would be expected by star-formation, considering the SFR of the galaxy. We used the $L_{2-10}-SFR$ relations of \citet{Ranalli:2003yb} and  \citet{Lehmer:2010hj}, and found that all sources in which no AGN were identified by our analysis have luminosities consistent, or lower, than the value expected from their SFR (see Fig.\,\ref{fig:lxvssfr}). The large fraction of U/LIRGs located below the relation of \citet{Ranalli:2003yb} is consistent with the flattening reported by \citet{Torres-Alba:2018yj} for a large sample of U/LIRGs observed by {\it Chandra}, and is possibly associated with an increasing level of obscuration within the star-forming regions  \citep{Torres-Alba:2018yj}.

\begin{table*}
\begin{center}
\caption[]{(1) IRAS name and (2) counterparts, observed (3) 2--10\,keV and (4) 10--24\,keV luminosities, intrinsic (5) 2--10\,keV and (6) 10--24\,keV AGN luminosities, (7) line-of-sight column densities towards the AGN and (8) references. Luminosity upper limits are calculated based on the observed flux, and therefore could be significantly higher if the source is heavily obscured. A line-of-sight column density of $N_{\rm\,H}=10^{24}\rm\,cm^{-2}$ ($10^{25}\rm\,cm^{-2}$) would correspond to an increase in luminosity of $\Delta [\log(L_{2-10}/\rm erg\,s^{-1})]=1.3$ ($2.8$)  $\Delta [\log(L_{10-24}/\rm erg\,s^{-1})]=0.4$ ($1.9$) in the 2--10\,keV and 10--24\,keV bands, respectively. The 2--10\,keV AGN luminosity upper limit was extrapolated (assuming a power-law with $\Gamma=1.8$) from the upper limit on the 10--24\,keV luminosity inferred by {\it NuSTAR}.}
\label{tab:Xrayprop}
\begin{tabular}{llccccccc}
\noalign{\smallskip}
\noalign{\smallskip}
\hline \hline \noalign{\smallskip}
   &   & \multicolumn{2}{c}{Observed} & \phantom{} &\multicolumn{3}{c}{Intrinsic (AGN)} \\
\multicolumn{1}{l}{  } & 
\multicolumn{1}{l}{ } & 
\multicolumn{2}{l}{} & \phantom{} &
\multicolumn{2}{l}{}  & \\ \cline{3-4} \cline{6-8} 
 \noalign{\smallskip}
 (1) & (2) & (3) & (4)& \phantom{}  &(5) &(6) &(7) &(8)  \\
\noalign{\smallskip}
IRAS name & Source  & $\log L_{2-10}$ & $\log L_{10-24}$ & \phantom{}  & $\log L_{2-10}$ & $\log L_{10-24}$ & $\log N_{\rm H}$ & Reference  \\
\noalign{\smallskip}
  &    & [$\rm erg\,s^{-1}$] & [$\rm erg\,s^{-1}$] & \phantom{}  & [$\rm erg\,s^{-1}$] & [$\rm erg\,s^{-1}$] & [$\rm cm^{-2}$] &  \\
\noalign{\smallskip}
\hline \noalign{\smallskip}
%
\noalign{\smallskip}
F00085$-$1223  &   NGC\,34                   &     41.41         &     	41.63	& \phantom{}      &  42.05 &	41.82 & 23.72 [23.62 -- 23.81]  	&          \cite{Ricci:2017aa}	\\        
\noalign{\smallskip}
\noalign{\smallskip}
F00163$-$1039  &   MCG$-$02$-$01$-$051                    &     40.95         &   $<40.52$    & \phantom{}           & $<40.68$	 &	$<40.52$	 & --  &            \cite{Ricci:2017aa} \\       
\noalign{\smallskip}
 &   MCG$-$02$-$01$-$052		  	  &     40.21         &  $<40.89$   & \phantom{}       &	$<41.05$  &	$<40.89$	 & -- &            \cite{Ricci:2017aa} \\    
\noalign{\smallskip}
\noalign{\smallskip}
F00344$-$3349  & ESO\,350$-$IG038      &     41.10         &  $<40.44$   & \phantom{}      &	 $<40.60$	& $<40.44$	 &       --     	 & This work  \\   
\noalign{\smallskip}
\noalign{\smallskip}
 F00506+7248 &   MCG+12$-$02$-$001             &    40.66          &  $<40.50$    & \phantom{}     &		$<40.66$ 	&	$<40.50$	&    --         & \cite{Ricci:2017aa} \\          
\noalign{\smallskip}
\noalign{\smallskip}
F01053$-$1746 & IC\,1623A/B     &     41.33         &   40.95     & \phantom{}    &	 $<41.11$	& 	$<40.95$ 	&       --      & This wok  \\   
\noalign{\smallskip}
\noalign{\smallskip}
F02069$-$1022  &   NGC\,833		  &    41.40          &   41.65       	& \phantom{} 	& 41.81	&	41.72 &      23.45 [23.40 -- 23.49]      & 	\cite{Oda:2018tr}\\    
\noalign{\smallskip}
   &   NGC\,835				   &      41.64        &   41.86      & \phantom{}    & 42.06	&	41.97&     23.63 [23.52 -- 23.76]        & \cite{Oda:2018tr}	 \\    
\noalign{\smallskip}	
\noalign{\smallskip}	
F02401$-$0013  &   NGC\,1068     &       41.19       &   41.39   & \phantom{}      &	 43.11	&	42.95  &        $\geq 24.99$    	 & \cite{Bauer:2015ve}  \\   
\noalign{\smallskip}
\noalign{\smallskip}	
F03117+4151   &  Mrk\,1073		  &     41.41         &   42.38  & \phantom{}      &	43.51 &    43.39      & 24.51 [24.34 -- 24.56]	 &  \citet{Yamada:2020nh}  \\   
\noalign{\smallskip}
\noalign{\smallskip}	
F03164+4119  &   NGC\,1275       	  &        44.58      &      43.50  & \phantom{}   & 43.22	&  43.06     &     21.68 [21.62 -- 21.78] 	 & \cite{Ricci:2017if}  \\   
\noalign{\smallskip}	
\noalign{\smallskip}	
F03316$-$3618 &    NGC\,1365	  &     41.71         &  41.79  & \phantom{}      & 42.00	 &    41.84   &     23.30 [23.28 -- 23.32]	 &  \cite{Lanz:2019nc}  \\   
\noalign{\smallskip}	
\noalign{\smallskip}	
F04454$-$4838 &ESO\,203$-$IG001  &	  $<41.38$ &	$<41.22$                   & \phantom{}   & $<41.38$ &	$<41.22$	 &     --       	 &  This work  \\   
\noalign{\smallskip}
\noalign{\smallskip}	
F05054+1718  &   CGCG 468$-$002E		  &      --        &  --     & \phantom{}     &	 --	& --	&       --      & \cite{Ricci:2017aa}	 \\    
\noalign{\smallskip}
  &   CGCG 468$-$002W		    &      42.83        &      42.80    	& \phantom{} & 42.84	&	42.80	&     22.18 [22.15 -- 22.20]        &  \cite{Ricci:2017aa} \\    
\noalign{\smallskip}	
\noalign{\smallskip}
F05189$-$2524 &  IRAS\,05189$-$2524		  &      43.10        &   43.30   & \phantom{}     & 43.57	&	43.02 &    23.10 [23.08 -- 23.14]        & \cite{Teng:2015vn} 	 \\    
\noalign{\smallskip}	
\noalign{\smallskip}	
F06076$-$2139   & South    &       41.36       & 41.90      & \phantom{}    &	42.34 & 42.18	&    23.79 [23.66 -- 23.93]         & \cite{Privon:2020lf}   \\   
\noalign{\smallskip}
   & North    &     40.83         &   --    & \phantom{}    &	-- &  --	&  --       & \cite{Privon:2020lf}   \\   
\noalign{\smallskip}	
\noalign{\smallskip}	
07251$-$0248	&   			  &     40.09          &    $< 41.87$    & \phantom{}   & 	$<42.03$ &	$<41.87$  &      --      & This work	 \\    
\noalign{\smallskip}	
\noalign{\smallskip}	
F08354+2555 & NGC\,2623   	  &     40.90         &   40.87   & \phantom{}     & 41.04 &	40.87 &         22.85 [22.63 -- 23.08]    	 &  This work \\   
\noalign{\smallskip}	
\noalign{\smallskip}	
F08520$-$6850  &   ESO\,060$-$IG16 	    &      41.83        &    41.93   & \phantom{}    & 42.11	& 41.94 	&        23.18 [22.95 -- 23.40]     	 & This work   \\   
\noalign{\smallskip}
\noalign{\smallskip}
F08572+3915  &   IRAS\,08572+3915			  &      41.06        &    $<41.13$    & \phantom{}   & 	$<41.29$ &	$<41.13$  &      --      & This work	 \\    
\noalign{\smallskip}
\noalign{\smallskip}
IRAS\,F09111$-$1007	&   	 W		  &        41.02       &    $< 41.36$    & \phantom{}   & 	$<41.52$ &	$<41.36$  &      --      & This work	 \\    
\noalign{\smallskip}
IRAS\,F09111$-$1007	&   	E		  &    41.16           &    $< 41.36$    & \phantom{}   & 	$<41.52$ &	$<41.36$  &      --      & This work	 \\    
\noalign{\smallskip}
\noalign{\smallskip}
F09320+6134  &   UGC\,05101					  &   41.77           &   42.83   & \phantom{}     &	43.43	&	43.23 &   24.11 [23.98 -- 24.21]         & \cite{Oda:2017bf}	 \\    
\noalign{\smallskip}
\noalign{\smallskip}
F09333+4841  &   MCG+08$-$18$-$013     &      40.80        &    $<40.44$      & \phantom{}   &	 $<40.60$  &          $<40.44$ & --  & \cite{Ricci:2017aa}	   \\     
\noalign{\smallskip}
 			 &   MCG+08$-$18$-$012	  	  &           $<40.64$ 	&	  $<40.48$      & \phantom{}     &	 $<40.64$ 	&	  $<40.48$      & --   & \cite{Ricci:2017aa}	 \\    
\noalign{\smallskip}
\noalign{\smallskip}
 F10015$-$0614 &   NGC\,3110                   	  &      40.59        &   $<40.61$     & \phantom{}   &	  $<40.77$ 	&	$<40.61$	 &  --          &  \cite{Ricci:2017aa} \\                 
\noalign{\smallskip}
 			 & MCG$-$01$-$26$-$013	  &    39.84          &    $<40.27$     & \phantom{}  	&  $<40.43$   &	 $<40.27$	&     --        & \cite{Ricci:2017aa} \\    	 
\noalign{\smallskip}	
\noalign{\smallskip}	
F10038$-$3338   &  IRAS\,F10038$-$3338        &       40.86        &   $ <40.81$        & \phantom{} 	& $<40.97$ 	&  $ <40.81$ 	 &    --    	 &  This work \\   
\noalign{\smallskip}
\noalign{\smallskip}
F10257$-$4339  &   NGC\,3256			  &      40.92        &  40.23     & \phantom{}    &   $<40.39$ 	&	$<40.23$ & --           & \cite{Lehmer:2015ys}	 \\    
\noalign{\smallskip}
\noalign{\smallskip}
F10565+2448  &   IRAS\,10565+2448		  &     41.31         & $<41.09$      & \phantom{}     & $<41.25$ &	    $<41.09$    & --    &  This work 	 \\    
%
\noalign{\smallskip}
 F11257+5850 &   Arp\,299W	  &       41.22       &    41.30  & \phantom{}     & 43.18		&	42.98 &     24.54 [24.52 -- NC]       & \cite{Ptak:2015nx}	 \\    
 \noalign{\smallskip}
   &   Arp\,299E					  &   41.01           &    --    & \phantom{}   	&	--	&	-- &      --      & \cite{Ptak:2015nx}	 \\     
\noalign{\smallskip}
\noalign{\smallskip}
\hline
\noalign{\smallskip}
\end{tabular}
\end{center}
\end{table*}

\setcounter{table}{1}
\begin{table*}
\begin{center}
\caption[]{Continued.}
\begin{tabular}{llccccccc}
\noalign{\smallskip}
\noalign{\smallskip}
\hline \hline \noalign{\smallskip}
   &   & \multicolumn{2}{c}{Observed} & \phantom{} &\multicolumn{3}{c}{Intrinsic (AGN)} \\
\multicolumn{1}{l}{  } & 
\multicolumn{1}{l}{ } & 
\multicolumn{2}{l}{} & \phantom{} &
\multicolumn{2}{l}{}  & \\ \cline{3-4} \cline{6-8} 
 \noalign{\smallskip}
 (1) & (2) & (3) & (4)& \phantom{}  &(5) &(6) &(7) &(8)  \\
\noalign{\smallskip}
IRAS name & Source  & $\log L_{2-10}$ & $\log L_{10-24}$ & \phantom{} & $\log L_{2-10}$ & $\log L_{10-24}$ & $\log N_{\rm H}$ & Reference  \\
\noalign{\smallskip}
  &    & [$\rm erg\,s^{-1}$] & [$\rm erg\,s^{-1}$] & \phantom{}  & [$\rm erg\,s^{-1}$] & [$\rm erg\,s^{-1}$] & [$\rm cm^{-2}$] &  \\
\noalign{\smallskip}
\hline \noalign{\smallskip}
\noalign{\smallskip}
\noalign{\smallskip}
 F12043$-$3140 &   ESO\,440$-$IG058N               &    40.41          &    $<40.79$     & \phantom{}      & $<40.95$ 	&	$<40.79$	 &    --        & \cite{Ricci:2017aa}  \\              
\noalign{\smallskip}
 			  &   ESO\,440$-$IG058S            &        40.22      &    $<40.87$      & \phantom{}      &	$<41.03$	 &	$<40.87$	&       --      &  \cite{Ricci:2017aa} \\           
\noalign{\smallskip}	
\noalign{\smallskip}	
F12112+0305   &   	  &      41.41        &    $<41.73$   & \phantom{}     &  $<41.89$  &	$<41.73$ &        --     	 &  This work \\   
\noalign{\smallskip}	
\noalign{\smallskip}	
F12243$-$0036  &  NGC\,4418 	& 	39.35   &    $<39.49$       & \phantom{}    &          $<39.65$  	& $<39.49$	 &       --     	 & This work   \\   
\noalign{\smallskip}
\noalign{\smallskip}
 F12540+5708 &   Mrk\,231								  &      42.58        &       42.67   	  & \phantom{} 	& 42.66	&	42.71	&      23.16 [23.08 -- 23.25]       &   \cite{Teng:2014oq}  \\    
\noalign{\smallskip}
\noalign{\smallskip}
 F12590+2934 &   NGC\,4922N                             &         41.07     &     41.55    & \phantom{}   	&	43.05	&	42.73	 & 25.10 [24.63 -- NC] & \cite{Ricci:2017aa} \\         
\noalign{\smallskip}
 &   NGC\,4922S                             &       38.81       &     --    & \phantom{}   	&	--	&	--	 & -- & \cite{Ricci:2017aa} \\         
\noalign{\smallskip}
\noalign{\smallskip}
13120$-$5453  &   IRAS\,13120$-$5453				  &      41.61        &         41.47 		& \phantom{}   	 & 43.10	& 42.94	 & 24.50 [24.27 -- 24.74]	& \cite{Teng:2015vn} 	 \\    
\noalign{\smallskip}	
\noalign{\smallskip}	
F13126+2453  &   IC\,860              &      38.55        &       $<39.60$    & \phantom{} 	& $<39.76$	&	$<39.60$ & --	 & This work   \\   
\noalign{\smallskip}	
\noalign{\smallskip}	
F13188+0036      & NGC\,5104        &    40.24          &   $<40.60$   & \phantom{}      &$<40.76$	&	$<40.60$  & --	 &  \cite{Privon:2020lf}   \\  
\noalign{\smallskip}	
\noalign{\smallskip}	
F13197$-$1627	&	 MCG$-$03$-$34$-$063						  &         $<40.13$     &  $<39.87$      & \phantom{}  	&  $<40.13$	  & $<39.87$ & -- &  \cite{Ricci:2017aa} \\ 
\noalign{\smallskip}	
	&	MCG$-$03$-$34$-$064 						    &       42.25       &   42.94	    & \phantom{}   	&	43.41  &	43.20 & 23.73 [23.72 -- 23.74] &  \cite{Ricci:2017aa} \\ 
\noalign{\smallskip}	
\noalign{\smallskip}	
F13229$-$2934   &  NGC\,5135	  &      41.20        &    42.06  & \phantom{}     &	43.35  &43.19 	& 24.80 [24.51 -- 25.00]	 & \cite{Yamada:2020nh}  \\  
\noalign{\smallskip}	
\noalign{\smallskip}	
F13362+4831 &	NGC\,5256-NE    &      41.54        &    41.42   & \phantom{}    &	 41.60	&	41.44  &	22.83 [22.48 -- 23.03]  & \cite{Iwasawa:2020ne}   \\  
\noalign{\smallskip}	
 &	NGC\,5256-SW   &     41.08         &   41.73     & \phantom{}   &	 43.13	&	42.97  &	$>24.30$ & \cite{Iwasawa:2020ne}   \\  
\noalign{\smallskip}
\noalign{\smallskip}
F13428+5608  &   Mrk\,273							  &    42.42          &      42.61    	& \phantom{}    &	42.93 &	42.96 &	23.64	[23.58 -- 23.73] &  \cite{Teng:2015vn}  \\    
\noalign{\smallskip}	
\noalign{\smallskip}	
F14348$-$1447 &   NE &	41.13	  &      $<41.85$        &        \phantom{}    & $<42.01$	&	 $<41.85$	 &  -- &  This work \\  
\noalign{\smallskip}	
 & SW  &	41.54	  &      $<41.85$        &        \phantom{}    & $<42.01$	&	 $<41.85$	 &  -- &  This work \\  
\noalign{\smallskip}
\noalign{\smallskip}
F14378$-$3651  &   IRAS\,14378$-$3651						    &      41.34        &    $<41.71$  & \phantom{}     	 &	 $<41.87$    &	$<41.71$	 & --  & This work  \\    
\noalign{\smallskip}
\noalign{\smallskip}
F14544$-$4255  &   IC\,4518A				  &       42.40       &     42.70   & \phantom{}   	  & 42.85	&	42.75 & 23.38 [23.34 -- 23.41] & \cite{Ricci:2017aa}	 \\     
\noalign{\smallskip}
 			 &   IC\,4518B				  &      40.57        &   --    & \phantom{}    	  & 41.09	&	40.89 & 23.51  [23.26 -- 23.86]  & \cite{Ricci:2017aa}	 \\    
\noalign{\smallskip}	
\noalign{\smallskip}	
F15250+3608  &    &		40.52	  &        $<41.66$      &      \phantom{}     &$<41.82$	 & $<41.66$ & -- &  This work  \\  
\noalign{\smallskip}
\noalign{\smallskip}
F15327+2340  &   Arp\,220W								   &       40.81       &    40.89     & \phantom{}   	&	$\gtrsim 42.92$	& $\gtrsim 42.72$ & $>24.72$	 & \cite{Teng:2015vn}	 \\    
\noalign{\smallskip}
\noalign{\smallskip}
F16504+0228  &   NGC\,6240 -- North							  &       42.03       &  42.42    & \phantom{}     	& 43.30	& 43.17 & 24.19 [24.09 -- 24.36]	 & \cite{Puccetti:2016lu}	 \\    	
\noalign{\smallskip}
 			 &   NGC\,6240 -- South									  &       42.38       &   42.86      & \phantom{}  	 & 	43.72 &	43.58  & 24.17 [24.11 -- 24.23] & \cite{Puccetti:2016lu}	 \\    	
\noalign{\smallskip}
\noalign{\smallskip}
 F16577+5900 &   NGC\,6286                         &    40.81          &      41.46     & \phantom{}     & 41.98	&	41.78	 & 24.05 [23.85 -- 24.34] & \cite{Ricci:2016zr}  \\      
\noalign{\smallskip}
 			 &   NGC\,6285                   &    40.22          &     $<40.35$            & \phantom{}       &	 $<40.51$	&	$<40.35$  & --  &  \cite{Ricci:2016zr}	 \\    
\noalign{\smallskip}
\noalign{\smallskip}
F17138$-$1017  &   IRAS\,F17138$-$1017         &       41.00       &    $<41.20 $    & \phantom{}       &	 $<41.36$	&	$<41.20 $	 & -- &  \cite{Ricci:2017aa}	 \\    
\noalign{\smallskip}
\noalign{\smallskip}
F17207$-$0014  &   IRAS\,F17207$-$0014        &       41.41       &    $<41.37$        & \phantom{}   	&	 $<41.53$	&	$<41.37$	 & -- &  This work \\    
\noalign{\smallskip}
\noalign{\smallskip}
F18293$-$3413  &   IRAS\,F18293$-$3413           &        41.02      &   $<40.60$       & \phantom{}      &	$<40.76$ 	&	$<40.60$   & --  & This work 	 \\                 
\noalign{\smallskip}
\noalign{\smallskip}
F19297$-$0406 &     &      41.74        &    $<42.01$      & \phantom{} 	 &	$<42.17$	&	$<42.01$ & --  & This work \\  
\noalign{\smallskip}
\noalign{\smallskip}
F20221$-$2458      & NGC\,6907&   40.18    &    $<40.22$    & \phantom{}       &          	 $<40.38$	&	$<40.22$  & --	 &  \cite{Privon:2020lf}  \\  
\noalign{\smallskip}
\noalign{\smallskip}
20264+2533  &   MCG $+$04$-$48$-$002 		  &    41.48          &     42.14     & \phantom{} 	 & 42.36	&	42.38 & 23.86 [23.79 -- 23.92]  & \cite{Ricci:2017aa}	 \\    		
\noalign{\smallskip}
 			 &   NGC\,6921				  &      41.28        &   42.22    & \phantom{}    & 42.97	&42.72	 &  24.15 [23.83 -- 24.40]  & \cite{Ricci:2017aa}	 \\  
\noalign{\smallskip}	
\noalign{\smallskip}	
F20550+1655  & CGCG\,448$-$020W   	  &     39.46         &  $<41.25$     & \phantom{}       &	 $<41.41$	& $<41.25$	 & --	 & This work   \\  
\noalign{\smallskip}	
  & CGCG\,448$-$020E   	  &     41.05         &  $<41.25$     & \phantom{}       &	 $<41.41$	& $<41.25$	 & --	 & This work   \\  
\noalign{\smallskip}
\noalign{\smallskip}
\hline
\noalign{\smallskip}
\end{tabular}
\end{center}
\end{table*} 

\setcounter{table}{1}
\begin{table*}
\begin{center}
\caption[]{Continued.}
\begin{tabular}{llccccccc}
\noalign{\smallskip}
\noalign{\smallskip}
\hline \hline \noalign{\smallskip}
   &   & \multicolumn{2}{c}{Observed} & \phantom{} &\multicolumn{3}{c}{Intrinsic (AGN)} \\
\multicolumn{1}{l}{  } & 
\multicolumn{1}{l}{ } & 
\multicolumn{2}{l}{} & \phantom{} &
\multicolumn{2}{l}{}  & \\ \cline{3-4} \cline{6-8} 
 \noalign{\smallskip}
 (1) & (2) & (3) & (4)& \phantom{}  &(5) &(6) &(7) &(8)  \\
\noalign{\smallskip}
IRAS name & Source  & $\log L_{2-10}$ & $\log L_{10-24}$ & \phantom{} & $\log L_{2-10}$ & $\log L_{10-24}$ & $\log N_{\rm H}$ & Reference  \\
\noalign{\smallskip}
  &    & [$\rm erg\,s^{-1}$] & [$\rm erg\,s^{-1}$] & \phantom{}  & [$\rm erg\,s^{-1}$] & [$\rm erg\,s^{-1}$] & [$\rm cm^{-2}$] &  \\
\noalign{\smallskip}
\hline \noalign{\smallskip}
\noalign{\smallskip}	
\noalign{\smallskip}	
F20551$-$4250  &   ESO\,286$-$IG19      &         41.65     &   $<41.50$      & \phantom{}     &	 	$<41.66$	&	$<41.50$  & --	 & This work  \\  
\noalign{\smallskip}
\noalign{\smallskip}
F21453$-$3511  &   NGC\,7130	  &       41.11       &    41.88      	& \phantom{} 	& 43.05	&	42.62 & 24.61	[24.50 -- 24.66]   & \cite{Ricci:2017aa}	 \\    
\noalign{\smallskip}
\noalign{\smallskip}	
F23007+0836  &   NGC\,7469				  &      43.24        &      43.10   & \phantom{}  	 &	43.36  & 43.14  & 19.78	[19.60 --19.90]	 & \cite{Ricci:2017aa}	 \\    
\noalign{\smallskip}
 			  &   IC\,5283				  &        --      &     --     		& \phantom{}  &  --  &	--   &   & \cite{Ricci:2017aa}	 \\    
\noalign{\smallskip}	
\noalign{\smallskip}	
F23128$-$5919  &ESO\,148$-$IG002     &        41.94      &    $ <41.23$      & \phantom{}   &	 $<41.39$	&	$ <41.23$ & --  &	 This work   \\  
\noalign{\smallskip}
\noalign{\smallskip}
F23157+0618     & NGC\,7591    &     39.97         &   $<40.26$    & \phantom{}    &	 $<40.42$	& $<40.26$	  &	--  &  \cite{Privon:2020lf}   \\  
\noalign{\smallskip}
\noalign{\smallskip}
F23254+0830  &   NGC\,7674		  &     42.14         &    42.52     & \phantom{}  	&  43.60	&	43.44	 & $ >24.48$ & \cite{Gandhi:2017da}  \\    
\noalign{\smallskip}
 			  &   NGC\,7674A		  &   --           &   --       & \phantom{} 				&  --	&	 --	 &   & \cite{Gandhi:2017da}  \\    
\noalign{\smallskip}
\noalign{\smallskip}
 23262+0314 &   NGC\,7679			  &      41.60        &        41.50  		& \phantom{}   &	41.60  &	41.50  & $<20.30$   & \cite{Ricci:2017aa}	 \\    
\noalign{\smallskip}
                           &   NGC\,7682			  &     41.28         &       42.28   	& \phantom{} 	  &  43.70	&43.30	 &  24.39 [23.99 -- 24.48] & \cite{Ricci:2017aa}	 \\    
\noalign{\smallskip}
\noalign{\smallskip}
F23365+3604   &        &        41.48      &   $<41.76$        & \phantom{} 	 &	  $<41.92$  &	$<41.76$   &   --  &  This work	 \\   
\noalign{\smallskip}
\noalign{\smallskip}
\hline
\noalign{\smallskip}
\end{tabular}
\end{center}
\end{table*}

\begin{figure}
\centering
\includegraphics[width=8.8cm]{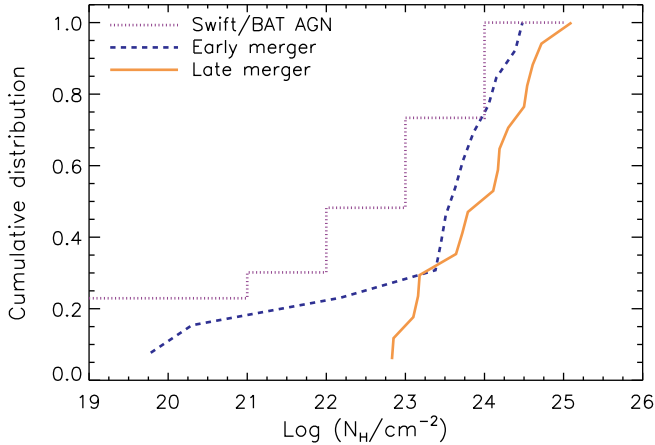}
  \caption{Column density cumulative distribution function for the AGN in our sample found in early (dashed blue line) and late mergers (continuous orange line). AGN found in galaxies undergoing the final phases of a merger are more obscured than those in the early mergers, and than hard X-ray selected nearby AGN (dotted purple line; \citealp{Ricci:2015fk,Ricci:2017if}).}
\label{fig:NHcdfvsStages}
\end{figure}

\section{Obscuration and X-ray properties of AGN in mergers}\label{sec:agnobscuration}

\subsection{X-ray AGN}

With the goal of understanding the evolution of AGN obscuration in the final phase of a galaxy merger, we report here the results obtained by studying 60 U/LIRGs from GOALS, effectively doubling the sample observed by NuSTAR presented by \citeauthor{Ricci:2017aa} (\citeyear{Ricci:2017aa}; see Table\,\ref{tab:Xrayprop}). This was done including, besides the 23 sources presented here, the 30 objects reported in \citet{Ricci:2017aa}, and several sources reported in recent literature (e.g., \citealp{Privon:2020lf,Yamada:2020nh,Iwasawa:2020ne}).
Using X-ray spectroscopy, we identify a total of 35 AGN in these systems, five of which are associated with U/LIRGs in the N stage (i.e. showing no clear sign of interactions). In the following we will refer to objects which were not identified as AGN in the X-rays, and for which only an upper limit of the AGN X-ray luminosity is reported in Table\,\ref{tab:Xrayprop}, as {\it X-ray non-AGN}.

The overall fraction of CT AGN for the sample is 16/35 ($46\pm8\%$\footnote{Fractions are calculated following \citet{Cameron:2011yw}, and the uncertainties quoted represent the 16th/84th quantiles of a binomial distribution, obtained using the Bayesian approach outlined in  \citet{Cameron:2011yw}.}), while that of AGN in merging galaxies is 13/30 ($44^{+8}_{-9}\%$). This is significantly higher than what is inferred for the hard X-ray selected {\it Swift}/BAT sample overall ($27\pm4\%$, \citealp{Ricci:2015fk,Ricci:2017if}), which is mostly composed of AGN in isolated galaxies (\S\ref{sec:comparisonBAT}). 
The sample of AGN in U/LIRGs shows a larger fraction of both heavily obscured ($N_{\rm\,H}\geq 10^{23}\rm\,cm^{-2}$; $82_{-7}^{+5}\%$) and obscured ($N_{\rm\,H}\geq 10^{22}\rm\,cm^{-2}$; $90^{+4}_{-6}\%$) sources than the {\it Swift}/BAT sample ($52\pm4\%$ and $70\pm 5\%$, respectively). This confirms the idea that the typical environment of these AGN is different from that of AGN in isolated galaxies, and that the obscuring medium almost fully covers the accreting SMBHs.

\subsection{AGN obscuration in the final phases of the merger}

Dividing our sample into different merger stages, excluding the $N$ stage galaxies, we have 13 and 17 AGN in early and late-stage mergers, respectively. We find that 4/13 ($33\pm12\%$) AGN in early mergers (i.e. objects classified as being in $a$ or $b$ stages) are CT, a fraction in good agreement with what is found for the {\it Swift}/BAT sample. This shows that it takes time to build up the obscuration, since at the beginning of the merger the fraction of CT AGN is not significantly higher than the comparison sample.
A large fraction of these objects (10/13 or $74^{+11}_{-12}\%$) are heavily obscured, and almost all of them (11/13 or $81^{+9}_{-11}\%$) are obscured. AGN in the late phases of the merger process (i.e., having $c$ or $d$ stages) show a higher fraction of CT AGN (9/17 or $53^{+11}_{-12}\%$) than both hard X-ray selected AGN and AGN in early-stage mergers. This is consistent with what was found by \cite{Ricci:2017aa}, and \citet{Guainazzi:2021nc} using {\it XMM-Newton}, who found that $\sim 47\%$ of the objects in their SDSS optically-selected sample are CT. Most of the AGN in late-stage mergers are heavily obscured (15/17 or $85^{+7}_{-9}\%$), and all of them are obscured. The difference between early mergers, late mergers, and hard X-ray selected AGN is clearly illustrated in Fig.\,\ref{fig:NHcdfvsStages}. AGN in the final phases of the merger process are consistently more heavily obscured than hard X-ray selected AGN, and do not show the tail of objects ($\sim 30\%$) with $\log (N_{\rm H}/\rm cm^{-2})\leq 23$ found in early mergers. Interestingly, 3/5 of the U/LIRGs in the N stage are CT ($58^{+18}_{-19}\%$), and 4/5 ($74^{+14}_{-19}\%$) are heavily obscured. This could be related to the fact that several of these systems are post-mergers.

\begin{figure}
\centering
\includegraphics[width=8.8cm]{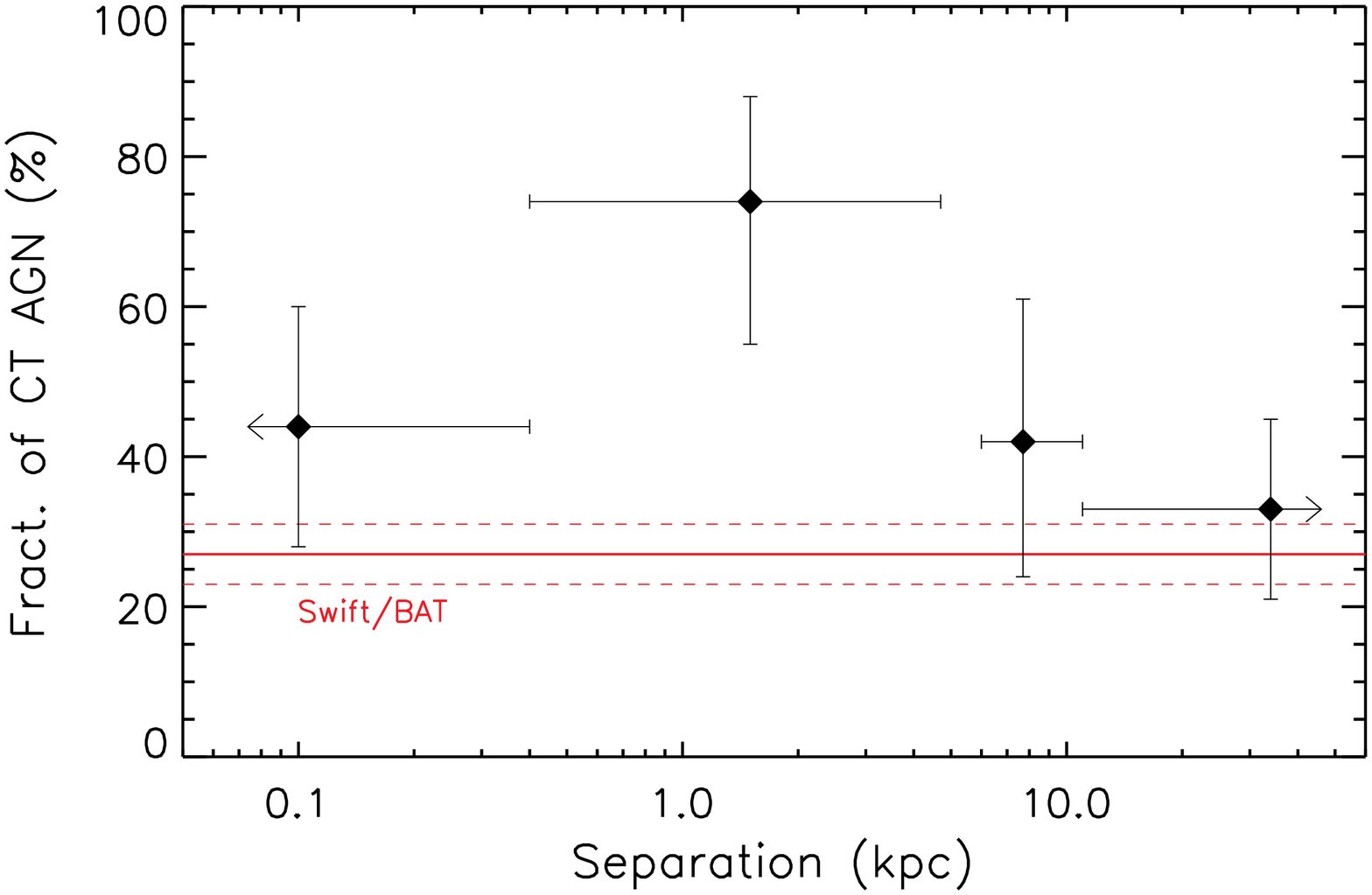}
\includegraphics[width=8.8cm]{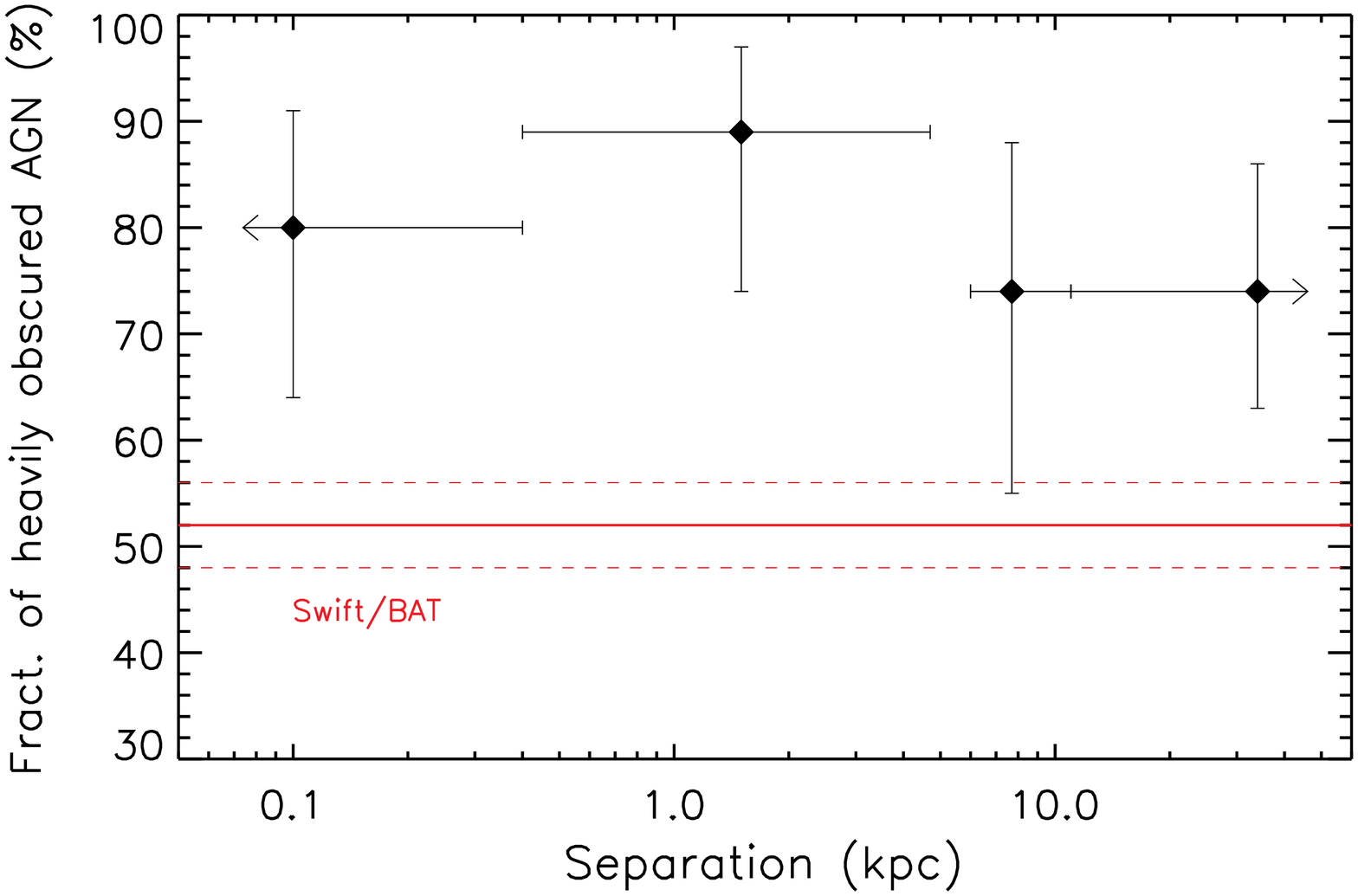}
  \caption{Fraction of CT ($N_{\rm H}\geq 10^{24}\rm\,cm^{-2}$; {\it top panel}) and heavily obscured ($N_{\rm H}\geq 10^{23}\rm\,cm^{-2}$; {\it bottom panel}) AGN in our sample versus the projected separation between the two nuclei. The fraction of heavily obscured sources appears to be consistently higher than {\it Swift}/BAT AGN (red continuous line; \citealp{Ricci:2015fk,Ricci:2017if}), and a tentative peak in the fraction of CT AGN is found at a separation of a few kpc.  Fractions are calculated following \citet{Cameron:2011yw}, and the uncertainties quoted represent the 16th/84th quantiles of a binomial distribution, obtained using the Bayesian approach outlined in  \citet{Cameron:2011yw}. }
\label{fig:fractionvsseparation}
\end{figure}

Recent simulations (e.g., \citealp{Blecha:2018gt,Kawaguchi:2020qb}) have shown that the most obscured phase during the merger would correspond to small separation between the two nuclei. We divided our sample based on the projected nuclear separation, down to the scales in which two nuclei could be resolved at the distance of our sources ($d_{\rm sep}\sim 0.4$\,kpc). In Fig.\,\ref{fig:fractionvsseparation} we show the fraction of CT (top panel) and heavily obscured (bottom panel) AGN versus the projected separation between the two galactic nuclei. We find that the CT fraction appears to peak when the two nuclei are at a projected distance of a few kpc ($74^{+14}_{-19}\%$; $d_{\rm sep}\sim 0.4-6$\,kpc), and that the fraction of heavily obscured sources is consistently higher than that found for {\it Swift}/BAT AGN, regardless of the projected nuclear separation. Similarly to the CT fraction, a peak in the median $N_{\rm H}$ [$(1.6\pm0.5)\times10^{24}\rm\,cm^{-2}$] is also observed when the merging galaxies are at a few kpc distance (Fig.\,\ref{fig:NHvsseparation}). The difference in the median column density with the {\it Swift}/BAT AGN sample (red continuous line) is particularly striking, with the X-ray observed U/LIRGs having a median $N_{\rm H}$ $\sim 1.5-2$ orders of magnitude larger. It should be noted that our U/LIRGs are frequently interacting or merging galaxy pairs, a process that increases the amount of gas within the central $\sim$kpc (e.g., \citealp{Di-Matteo:2007ly}). This could lead to the GOALS galaxies having additional obscuration on hundreds of pc to kpc scales, compared to host galaxy obscuration seen in AGN in non-merging systems. The median CO luminosity of GOALS galaxies ($2.6\times10^9\rm\,K\,km\,s^{-1}\,pc^2$, \citealp{Herrero-Illana:2019br}) is a factor of seven higher than the CO luminosity of host galaxies of BAT AGN ($4\times10^8\rm\,K\,km\,s^{-1}\,pc^2$, \citealp{Koss:2021ec}). Different molecular gas masses could affect the contribution of host galaxy obscuration, but not up to the CT level (e.g., \citealp{Buchner:2017iv}). However, due to potential variations in the CO$-H_{2}$ conversion factor (e.g., \citealp{Bolatto:2013ln}), it is unclear how different the total molecular gas masses are between the two samples.

\subsection{IR and X-ray luminosities of AGN in U/LIRGs}\label{sec:xraylum}

\begin{figure*}
\centering
\includegraphics[width=8.8cm]{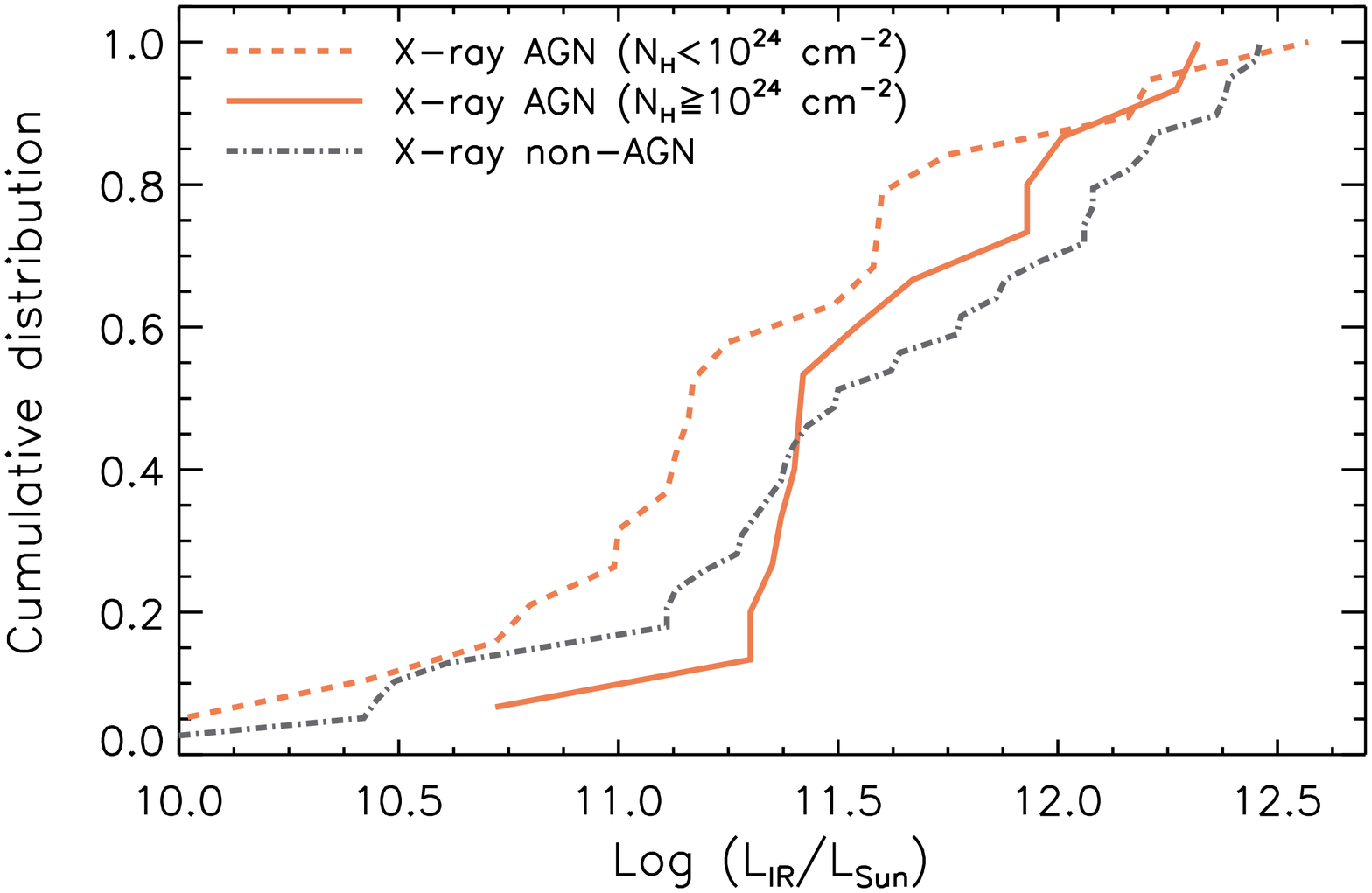}
\includegraphics[width=8.8cm]{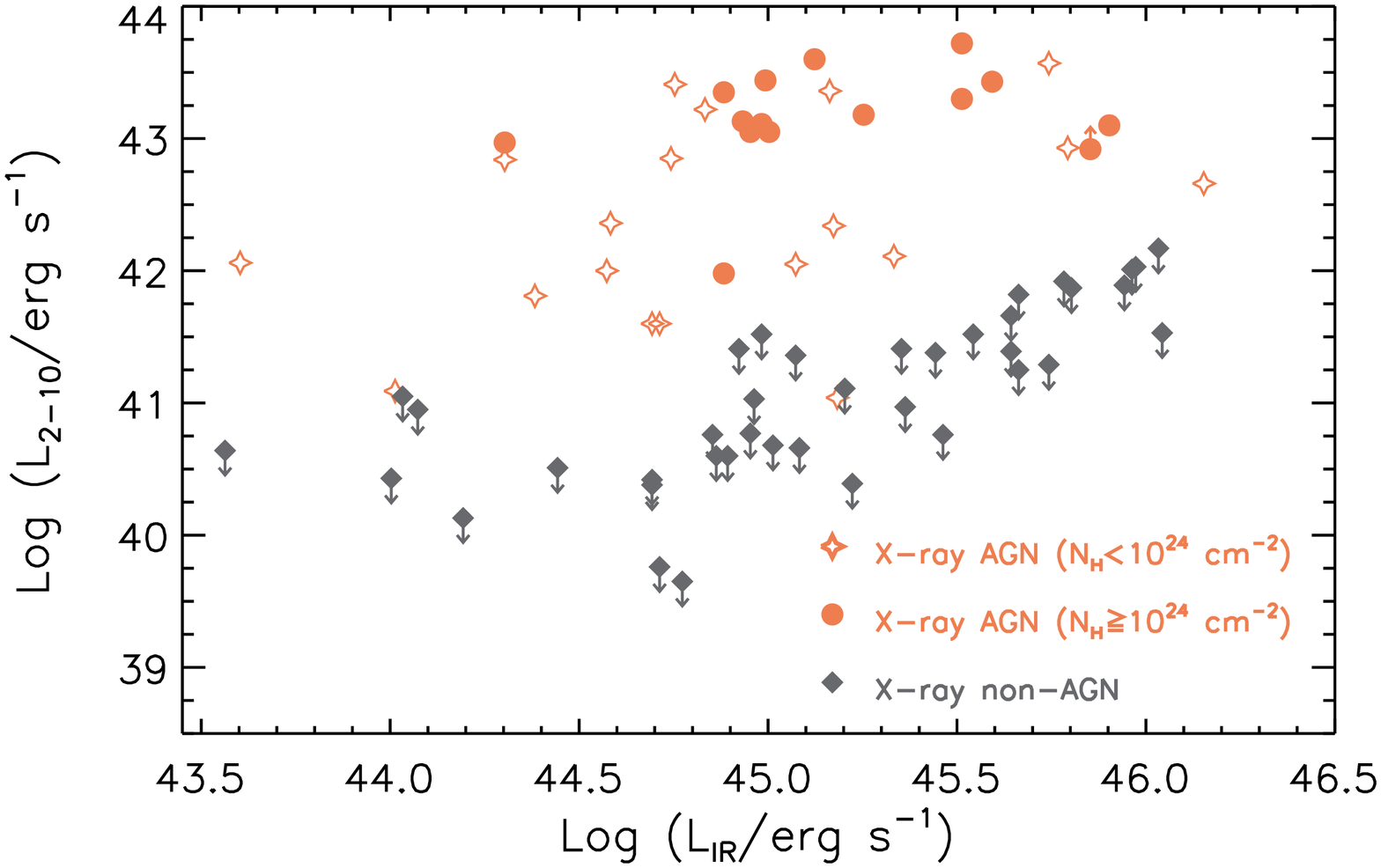}
  \caption{{\it Left panel:} Cumulative distribution function of the $8-1000\,\mu$m IR luminosities of X-ray non-AGN (grey dot-dashed line), Compton-thin (red dashed line) and CT (red continuous line) AGN, showing that CT AGN are typically found in systems that are more luminous in the IR. {\it Right panel:} Intrinsic 2--10\,keV luminosity versus $8-1000\,\mu$m luminosity of the X-ray detected Compton-thin (red empty stars) and CT (red filled circles) AGN in our sample. The plot also shows the upper limit on the 2--10\,keV luminosity of the sources for which an AGN was not identified in the X-rays (grey filled diamonds). These upper limits are calculated based on the observed flux, and therefore could be significantly higher if the source is heavily obscured. A line-of-sight column density of $N_{\rm\,H}=10^{24}\rm\,cm^{-2}$ ($10^{25}\rm\,cm^{-2}$) would correspond to an increase in luminosity of $\Delta [\log(L_{2-10}/\rm erg\,s^{-1})]=1.3$ ($2.8$). The figure illustrates that the CT AGN we identify typically have higher intrinsic X-ray luminosities than the Compton-thin AGN. }
\label{fig:LIRvsStages}
\end{figure*}

In the left panel of Fig.\,\ref{fig:LIRvsStages} we show the cumulative distribution of the IR luminosity of the X-ray detected AGN in our sample, divided into Compton-thin (blue dashed line) and CT (red continuous line). The CT AGN tend to have higher IR luminosities and, from performing a Kolmogorov-Smirnov (KS) test between the $L_{\rm IR}$ distributions of the two types of AGN, we find a p-value of 0.01. This indicates that the IR luminosities of CT and Compton-thin AGN are significantly different. Interestingly, CT AGN (red filled circles in the right panel of Fig.\,\ref{fig:LIRvsStages}) are mostly found to have higher intrinsic X-ray luminosities than Compton-thin sources (empty red stars), and only one of them is found to have $L_{2-10}\lesssim 10^{43}\rm\,erg\,s^{-1}$. A KS test between the two luminosity distributions results in a p-value of $4.6\times 10^{-4}$. While we cannot exclude that this is an evolutionary effect, where more obscured sources accrete more rapidly because they have more gas available in their surroundings, it is possible that this is related to a selection effect. In fact, even with our sensitive {\it NuSTAR} hard X-ray observations, because of the strong depletion of the X-ray flux at $N_{\rm H}> 10^{24}\rm\,cm^{-2}$, it would be difficult to detect a large number of low-luminosity heavily obscured AGN. This is particularly true if the heavily obscuring material covers most of the X-ray source, as suggested by the very large fraction of heavily obscured sources (see also \citealp{Ricci:2017aa}), which would lead to a small fraction of the X-ray radiation reprocessed by the circumnuclear environment being able to escape the system.

\subsection{Constraints on obscuration from IR-identified AGN}\label{sec:xrayobs_nondet}

\begin{figure}
\centering
\includegraphics[width=8.8cm]{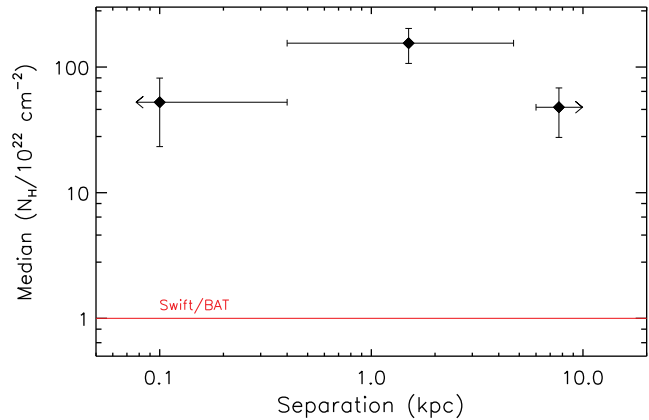}
  \caption{Median column density versus the separation between the two nuclei (Tables\,\ref{tab:sample} and \ref{tab:Xrayprop}). AGN in U/LIRGs tend to have significantly higher column densities than AGN in nearby hard X-ray selected AGN (red continuous line; \citealp{Ricci:2015fk,Ricci:2017if}). Similarly to what was found for the fraction of CT AGN (top panel of Fig.\,\ref{fig:fractionvsseparation}), we find a tentative peak of the column density for a projected nuclear separation of a few kpc. We conservatively considered the lower limit on $N_{\rm H}$ for the objects for which this parameter could not be constrained.}
\label{fig:NHvsseparation}
\end{figure}

Multi-wavelength tracers of AGN activity can help discover heavily obscured accreting SMBHs that cannot be identified in the X-rays. In a companion paper (Ricci et al. in prep.) we discuss in more detail these proxies of AGN activity for our sample of U/LIRGs, comparing them with the AGN X-ray emission. Considering [Ne\,V] as a good tracer of AGN activity (e.g., \citealp{Weedman:2005cr,Armus:2006yt,Goulding:2009dq,Petric:2011zt}), we can identify only one accreting SMBH (in the late-stage merger IRAS\,F23128$-$5919) that is not an X-ray detected AGN. Assuming that the [Ne\,V] 14.32$\mu$m emission is entirely due to the AGN, and invoking the [Ne\,V]/X-ray correlation from \cite{Satyapal:2007nx}, we would expect this source to have a bolometric AGN luminosity of $6\times10^{44}\rm\,erg\,s^{-1}$. Using a 2--10\,keV bolometric correction of $\kappa_{\rm X}=20$ (e.g., \citealp{Vasudevan:2007fk}), the source would need to be obscured by $N_{\rm H}\gtrsim 1.9\times10^{24}\rm\,cm^{-2}$ to have a 2--10\,keV AGN luminosity consistent with the upper limit inferred by our study (Table\,\ref{tab:Xrayprop}), which is reasonable given the $N_{\rm H}$ distribution of the other AGN.

Considering MIR photometry, and assuming a $W1-W2>0.8$ threshold for AGN activity \citep{Stern:2012fk}, we find 13 sources that are classified as AGN with {\it WISE} (including IRAS\,F23128$-$5919), but were not identified by our broad-band X-ray analysis. Of these, 11 are in the final stages ($c$ and $d$) of the merger process, one is not a merger (N), and another is stage $b$. This would be even further evidence for the CT nature of most late stage mergers. We use the method outlined by \citeauthor{Pfeifle:2021fx} (\citeyear{Pfeifle:2021fx}; see their Eq.\,2), assuming that the 12$\mu$m emission is dominated by the AGN in these sources, to get constraints on the $N_{\rm H}$ needed for these sources to remain undetected by {\it NuSTAR}. We find that all the lower limits on the column density are above $\sim 3\times 10^{24}\rm\,cm^{-2}$. Including these lower limits to our sample, a total of $69^{+8}_{-9}\%$ of the AGN in the final stages of mergers would be CT. Dividing this sample according to the nuclear separation, we find that the peak of the CT fraction would again be found at $\sim 0.4-6$\,kpc ($85^{+8}_{-12}\%$), while $62^{+13}_{-14}\%$ of the AGN in merging galaxies that show a single galactic nucleus would be CT. It should be noted that, considering these candidate AGN, we are still able to detect with {\it NuSTAR} the accreting SMBHs that are contributing to most of the overall IR emission. Using the 12$\mu$m AGN emission for the {\it WISE}-selected candidate AGN that were not detected in the X-rays, considering a typical AGN IR spectrum \citep{Stalevski:2012ma,Stalevski:2016yf}, we find that accreting SMBHs would in fact contribute at most $\sim 40$\% to the IR flux.

\subsection{The evolution of obscuration in U/LIRGs}

This work and recent X-ray studies (e.g., \citealp{Ricci:2017aa}) show that the obscuration properties of AGN in U/LIRGs are very different from those of AGN in isolated galaxies. In particular, AGN in late mergers are fully embedded in gas with $N_{\rm H}\geq 10^{23}\rm\,cm^{-2}$. The most extreme of these sources could be associated with Compact Obscured Nuclei (or CONs, \citealp{Aalto:2015ni,Aalto:2019yr,Falstad:2021wf}), galaxies which show strong and compact vibrationally-excited HCN from their nuclear regions. This emission is created by a strong 14$\mu$m continuum, which could be due to strong emission from a heavily obscured AGN. The presence of obscuring material with a very high covering factor around AGN in galaxies undergoing the final phases of a merger has also been confirmed by a recent study focussing on the [O\,IV]\,25.89$\mu$m line. \citet{Yamada:2019wm} found that the ratio between the [O\,IV] and the 12$\mu$m AGN luminosity decreases as the merger progresses, which suggests that the covering factor of the material tends to be larger in late-stage mergers. \citet{Ricci:2017ej} demonstrated that, due to the presence of dusty gas (e.g., \citealp{Fabian:2006lq,Fabian:2008hc}), radiation pressure can be very effective in reducing the covering factor of the obscuring material, by removing gas from the environment of nearby AGN already at low Eddington ratios (i.e., $\lambda_{\rm Edd}\sim 10^{-2}$; see also \citealp{Garcia-Burillo:2021tb}). This process might not be as effective in mergers, where the obscuring material might be located at 100s of parsecs from the accreting source (and therefore outside the sphere of influence of the SMBH). In these objects, the AGN would need to attain high luminosities (and considerably higher Eddington ratios) in order to remove the obscuring material \citep{Ricci:2017ej,Jun:2021xp}. This might happen in the final stages of the merger process, when the accretion rate of the SMBH is expected to reach very high levels (e.g., \citealp{Blecha:2018gt,Kawaguchi:2020qb}), and it could be the cause of the tentative decrease at $d_{\rm sep}\lesssim0.4$\,kpc we observe both in the fraction of CT AGN (top panel of Fig.\,\ref{fig:fractionvsseparation}) and in the median $N_{\rm H}$ (Fig.\,\ref{fig:NHvsseparation}). Alternatively, the decrease could be due to sources being more heavily obscured in the final phases of the merger process, which would lead us to detect preferentially the least obscured AGN.

\section{Summary and conclusions}\label{sec:summary}

In this work we have studied broad-band X-ray observations of 60 nearby U/LIRGs from the GOALS sample to understand the link between AGN obscuration and galaxy mergers. A total of 35 X-ray detected AGN are identified in these systems, 30 of which reside in merging galaxies. We find that:
 \begin{itemize}
 \item The U/LIRGs in our sample show a higher fraction of heavily obscured ($N_{\rm H}\geq 10^{23}\rm\,cm^{-2}$; $82^{+5}_{-7}\%$) and CT AGN ($N_{\rm H}\geq 10^{24}\rm\,cm^{-2}$; $46\pm 8\%$) than local hard X-ray selected AGN ($52\pm4\%$ and $27\pm4\%$, respectively; \citealp{Ricci:2015fk,Ricci:2017if}; see Fig.\,\ref{fig:NHcdfvsStages}). The median line-of-sight column density towards AGN in U/LIRGs is also $\sim 1.5-2$ orders of magnitude larger than that of hard X-ray selected AGN (Fig.\,\ref{fig:NHvsseparation}).
 \item Roughly half ($53^{+11}_{-12}\%$) of the AGN in galaxies undergoing the final stages of mergers are CT. This fraction of CT sources is higher than that found in AGN in early mergers ($33\pm12\%$) and in local hard X-ray selected AGN. Considering the X-ray non detections of objects which are identified as AGN in the IR (\S\ref{sec:xrayobs_nondet}), the fraction of CT AGN in late-stage mergers value would be higher ($69^{+8}_{-9}\%$).
 \item A tentative peak in the fraction of CT AGN is found at nuclear projected separations of $d_{\rm sep}\sim0.4-6$\,kpc ($74_{-19}^{+14}\%$; top panel of Fig. \ref{fig:fractionvsseparation}). The median line-of-sight column density is also found to peak  [$(1.6\pm0.5)\times10^{24}\rm\,cm^{-2}$] for a similar range of nuclear separations (Fig.\,\ref{fig:NHvsseparation}). Considering the X-ray non detection of objects that are identified as AGN in the IR, the CT fraction at $d_{\rm sep}\sim0.4-6$\,kpc would be $85^{+8}_{-12}\%$. The possible decrease at $d_{\rm sep}\lesssim 0.4$\,kpc in both the fraction of CT AGN and in the median $N_{\rm H}$ could be related to the effect of radiation pressure, or to the fact that sources are more heavily obscured in the final phases of the merger process, and therefore we would detect preferentially the least obscured AGN.
 \item The vast majority ($85^{+7}_{-9}\%$) of the AGN in late-stage mergers are heavily obscured. This fraction is consistent with that obtained for early mergers ($74^{+11}_{-12}\%$), while it is significantly higher than for local hard X-ray selected AGN.
 \item CT AGN typically have higher intrinsic (i.e. absorption-corrected) X-ray luminosities than less obscured sources. This could either be due to an evolutionary effect, with more obscured sources accreting more rapidly because they have more gas available in their surroundings, or to a selection effect. In the latter scenario our {\it NuSTAR} observations might be unable to detect a significant fraction of heavily obscured less luminous ($L_{2-10}\lesssim 10^{43}\rm\,erg\,s^{-1}$) AGN, while detecting most of the AGN that contribute significantly to the energetics of these U/LIRGs (Ricci et al. in prep.).
 \end{itemize}
 
Our work confirms the idea that the close environments of AGN in U/LIRGs undergoing the final stages of the merger process are different from those of AGN in isolated galaxies (e.g., \citealp{Ricci:2017aa}), with the former having an accreting source completely buried by obscuring material. We speculate that, due to the high density and large covering factor of the obscuring dust and gas, there might be an important fraction of lower luminosity ($L_{2-10}\lesssim 10^{43}\rm\,erg\,s^{-1}$) AGN that we are still missing in late mergers. Extremely sensitive hard X-ray telescopes, such as those on board the proposed missions {\it FORCE} \citep{Mori:2016dx,Nakazawa:2018uh} and {\it HEX-P} \citep{Madsen:2018az}, would be fundamental to shed light on the accretion properties of SMBHs in these nearby systems. The strong nuclear obscuration associated with AGN in mergers, combined with the increase of galaxies in mergers with redshift (e.g., \citealp{Le-Fevre:2000ch,Conselice:2009dk,Lotz:2011ie}), might contribute to the observed positive relation between the fraction of obscured sources and redshift (e.g., \citealp{La-Franca:2005th,Treister:2006hn,Ueda:2014uq,Buchner:2015hb}). {\it Athena} \citep{Nandra:2013xe} will be a fundamental tool to assess the role of mergers in the increase of the fraction of obscured AGN with redshift, shedding light on the properties of accreting SMBHs at $z\gtrsim 1$.

\appendix
\section{X-ray observations log}\label{sect:appendix_obslog}

In Table\,\ref{tab:obslog} we report the X-ray observations used in our study. Details on the data reduction can be found in \S\ref{sec:datareduction}.

\begin{table*}
\centering
\caption{X-ray observations log. The table reports the name of the {\it IRAS} source and of the counterparts (columns 1 and 2, respectively), as well the X-ray observatory used (3), the ID (4), date (5) and exposure (6) of the observation.}\label{tab:obslog}
\begin{center}
\begin{tabular}{llcccc}
\noalign{\smallskip}
\hline \hline \noalign{\smallskip}
\multicolumn{1}{c}{(1)}  & \multicolumn{1}{c}{(2)} & (3) & (4) & (5) & (6)   \\
\noalign{\smallskip}
{\it IRAS} name & \multicolumn{1}{c}{Source} & Observatory & Obs. ID& Date & Exposure (ks) \\
\noalign{\smallskip}
\hline \noalign{\smallskip}
\noalign{\smallskip}	
F00344$-$3349  & ESO\,350$-$IG038                              &   {\it NuSTAR}   & 60374008002      &   2018-01-15      &     22.6                \\      
  			 &                  				             &  {\it Chandra}    & 8175       &  2006-10-28        &         54.0           \\      
\noalign{\smallskip}	
\noalign{\smallskip}	
F01053$-$1746 & IC\,1623A \& IC\,1623B                            &   {\it NuSTAR}  &  50401001002	     &  2019-01-19        &   20.6        \\      
 			&                         & {\it Chandra}     &  7063     &  2005-10-20      &   59.4                  \\      
 			&                         &   {\it XMM-Newton}   &  0830440101     & 2019-01-10       &      22.6               \\      
\noalign{\smallskip}	
\noalign{\smallskip}	
F04454$-$4838 &ESO\,203$-$IG001                              & {\it NuSTAR}     & 60374001002      & 2018-05-25          &    21.1              \\      
  &                           & {\it Chandra}     & 7802      &    2008-01-17        &     15.0           \\      
\noalign{\smallskip}	
07251$-$0248	&     				                           &  {\it NuSTAR}    &     60667003002     &   2021-04-09    &       32.6           \\     
		&     				                           &  {\it Chandra}    &    7804     &  2006-12-01      &        15.6        \\     
\noalign{\smallskip}	
\noalign{\smallskip}	
F08354+2555 & NGC\,2623                              &   {\it NuSTAR}   &  60374010002     &    2018-05-24      &        38.7            \\      
&                          &   {\it  Chandra}   &    4059    &    2003-01-03     &     19.8                \\      
&                          &   {\it  XMM-Newton}   &    0025540301    &   2001-04-27       &           4.9         \\      
\noalign{\smallskip}	
\noalign{\smallskip}	
F08520$-$6850  &   ESO\,060$-$IG16  (NE \& SW)                                 &     {\it NuSTAR} & 60101053002      &   2015-12-01       &          41.8        \\      
   			&                                      &  {\it Chandra}    &  7888     &   2007-05-31       &      14.7            \\      
\noalign{\smallskip}	
\noalign{\smallskip}	
F08572+3915  &    NW \& SE					                           &  {\it NuSTAR}    &  50401004002     &   2019-04-04     &     211.3            \\      
 &  					                           &  {\it NuSTAR}    &   60001088002    &   2013-05-23     &       24.1          \\      
 &  					                           &  {\it Chandra}    &   6862    & 2006-01-26     &          15.1         \\      
 &  					                           &  {\it XMM-Newton}    &   0830420101    &   2019-04-05      &    63.5            \\      
 &  					                           &  {\it XMM-Newton}    &   0830420101    &   2019-04-07      &     63.5           \\      
\noalign{\smallskip}	
F09111$-$1007	&     				                           &  {\it NuSTAR}    & 60667007002        & 2021-05-08      &          30.7        \\     
		&     				                           &  {\it Chandra}    &    7806    &   2007-03-20     &       14.8           \\     
\noalign{\smallskip}	
F10038$-$3338   &                                  &  {\it NuSTAR}    &   60101055002    & 2016-01-14        &      53.3               \\      
   &                                &  {\it Chandra}    &   7807    &   2007-03-07       &     14.4                \\      
\noalign{\smallskip}	
\noalign{\smallskip}	
F10565+2448  &   IRAS\,10565+2448						                           &    {\it NuSTAR}     &    60001090002   &    2013-05-22      &     25.3              \\      
&   						                           &    {\it Chandra}     &    3952   &    2003-10-23      &        28.9           \\   
&   						                           &    {\it XMM-Newton}     &    0150320201   &    2003-06-17      &        22.4           \\   
\noalign{\smallskip}	
\noalign{\smallskip}	
F12112+0305           &                   &   {\it NuSTAR}   &  60374005002     &    2018-01-17       &     15.3             \\      
          &                  &  {\it Chandra}    &  4110     &     2003-04-15      &         10.0         \\      
          &                  & {\it XMM-Newton}     &   0081340801    &    2001-12-30      &    16.2              \\      
\noalign{\smallskip}	
\noalign{\smallskip}	
F12243$-$0036  &   NGC\,4418                                        &   {\it NuSTAR}   & 60101052002      &  2015-07-03       &       43.8            \\      
  &                                         &   {\it Chandra}   & 4060      &   	2003-03-10       &   19.8                \\      
\noalign{\smallskip}	
\noalign{\smallskip}	
F13126+2453  &   IC\,860                                  &  {\it NuSTAR}    &  60301024002     &   2018-02-01         &       72.2           \\      
  &    	                       &    {\it Chandra}  & 10400      &      2009-03-24      &           19.2       \\      
\noalign{\smallskip}	
\noalign{\smallskip}	
F14348$-$1447           &     F14348$-$1447 (NE \& SW)               &       {\it NuSTAR}     &    60374004002       &   2018-01-27        &  21.0      \\      
                  &           &        {\it Chandra}    &    6861       &  2006-03-12      &      14.7      \\      
                  &           &         {\it XMM-Newton}    &    0081341401       &   2002-07-29      &     13.5    \\      
\noalign{\smallskip}	
\noalign{\smallskip}	
F14378$-$3651  &   IRAS\,14378$-$3651						                           &  {\it NuSTAR}    &  60001092002     &   2013-02-28        &    24.4               \\      
 &   				                           &  {\it Chandra}    & 7889      &     2007-06-25      &           13.9        \\   
\noalign{\smallskip}	
\noalign{\smallskip}	
F15250+3608                             &      &    {\it NuSTAR}   &   60374009002      &   2018-01-17     &    16.8          \\      
                             &      &    {\it Chandra}   &   4112      &  2003-08-27     &        9.8      \\      
                             &      &  {\it XMM-Newton}     &   0081341101      &   2002-02-22     &     14.9         \\      
\noalign{\smallskip}	
\noalign{\smallskip}	
F17207$-$0014  &   IRAS\,F17207$-$0014                                 &  {\it NuSTAR}    &     60667001002  &   2020-08-01      &       20.6              \\         
   &                                    &  {\it Chandra}    &  2035     &  2001-10-24        &           48.5          \\         
   &                                    &  {\it XMM-Newton}    &    0081340601   &    2002-02-19     &     12.2                \\         
\noalign{\smallskip}	
\noalign{\smallskip}	
F18293$-$3413  &                                 &  {\it NuSTAR}    &   60101077002    &  2016-02-20            &     21.2            \\      
  &                                &  {\it Chandra}    &    21379   &    2019-08-08          &       79.0          \\      
  &                                &   {\it XMM-Newton}   &    0670300701   &    2012-03-16         &       16.0          \\      
\noalign{\smallskip}
\noalign{\smallskip}
F19297$-$0406                             &      &  {\it NuSTAR}     &     60374007002  &  2018-03-03     &    20.0             \\      
                     			            &      &  {\it Chandra}     &   7890    &   2007-06-18     &         16.4       \\      
\noalign{\smallskip}
\hline
\noalign{\smallskip}
\end{tabular}
\end{center}
\end{table*}

\setcounter{table}{1}
\begin{table*}
	\centering
	\caption{Continued. }
\begin{center}
\begin{tabular}{llcccc}
\noalign{\smallskip}
\hline \hline \noalign{\smallskip}
\multicolumn{1}{c}{(1)}  & \multicolumn{1}{c}{(2)} & (3) & (4) & (5) & (6)   \\
\noalign{\smallskip}
{\it IRAS} name & \multicolumn{1}{c}{Source} & Observatory & Obs. ID& Date & Exposure (ks) \\
\noalign{\smallskip}
\hline \noalign{\smallskip}
\noalign{\smallskip}	
\noalign{\smallskip}	
F20550+1655  & CGCG\,448$-$020E \& CGCG\,448$-$020W      &  {\it NuSTAR}    &  60374002002     &  2018-03-28         &        24.9          \\      
  			&                            		&  {\it Chandra}    &   7818    &   2007-09-10        &       14.6           \\      
  			&                              		&  {\it XMM-Newton}    &   0670140101    &    2011-10-28       &      61.5            \\      
\noalign{\smallskip}	
\noalign{\smallskip}	
F20551$-$4250  &   ESO\,286$-$IG19                                   &   {\it NuSTAR}   &  60101054002     &     2015-07-30      &        42.6           \\      
  &                                     &   {\it Chandra}   &  2036     &     2001-10-31      &             44.9      \\      
  &                                     &   {\it XMM-Newton}   &  0081340401     &     2001-04-21    &      10.1             \\      
\noalign{\smallskip}	
\noalign{\smallskip}	
F23128$-$5919  &ESO\,148$-$IG002                            &  {\it NuSTAR}    & 60374006002      &    	2018-03-07        &          27.0       \\      
  &                             &  {\it Chandra}    &   2037    &     2001-09-30       &     49.3            \\      
  &                             &  {\it XMM-Newton}    &  0081340301     &   2002-11-19          &        8.4         \\      
\noalign{\smallskip}	
\noalign{\smallskip}	
F23365+3604	 &      &   {\it NuSTAR}       &     60667002002        &   2021-02-10           &  54.0   \\      
			 &      &  {\it Chandra}        &        4115     &       2003-02-03 &    10.1     \\      
\noalign{\smallskip}
\hline
\noalign{\smallskip}
\end{tabular}
\end{center}
\end{table*} 

\section{Results of the X-ray spectral analysis}\label{sect:appendix_spectralanalysis}

The results of the spectral fitting performed here are reported in Table\,\ref{tab:Xray_results1}. Details on the spectral fitting approach can be found in \S\ref{sec:xrayspectralmodelling}.

\begin{table*}
\centering
\caption{The table reports the values obtained from the X-ray spectral analysis of the sources of our sample. For each source we list (1) the IRAS name of the source, (2) the counterparts, (3) the column density of the X-ray emission associated with star-formation, (4) the temperature of the collisionally-ionised plasma, (5) the photon index of the soft X-ray emission due to X-ray binaries or to the scattered emission from the AGN, (6) the column density and (7) the photon index of the AGN, and (8) the value of the Cash  or $\chi^2$ statistics and the number of degrees of freedom (DOF). Objects in which both statistics were used to fit different spectra are reported as [C/$\chi^{2}$], and the value of the statistic (Stat) is the combination of the two. }\label{tab:Xray_results1}
\begin{center}
\begin{tabular}{llcccccc}
\noalign{\smallskip}
\hline \hline \noalign{\smallskip}
\multicolumn{1}{c}{(1)}  & \multicolumn{1}{c}{(2)} & (3) & (4) & (5) & (6) & (7) & (8)  \\
\noalign{\smallskip}
{\it IRAS} name & Source & 	$N_{\rm\,H}^{\rm SF}$ & 	kT & $\Gamma_{\rm\,bin.}$ & $N_{\rm\,H}$ & $\Gamma$  & Stat/DOF\\
\noalign{\smallskip}
 & 	&($10^{21}\rm\,cm^{-2}$) &  (keV) &   & ($10^{22}\rm\,cm^{-2}$) &    \\
\noalign{\smallskip}
\hline \noalign{\smallskip}
\noalign{\smallskip}
F00344$-$3349 & 	ESO\,350$-$IG038				& $0.9_{-0.3}^{+0.4}$ 	& $0.79^{+0.08}_{-0.09}$ &  $1.87_{-0.17}^{+0.18}$   &  --    & --  &  294/330   \\   
\noalign{\smallskip}
\noalign{\smallskip}
F01053$-$1746 &  IC\,1623A \& IC\,1623B &	 $0.8_{-0.1}^{+0.2}$ 	& $0.73^{+0.03}_{-0.04}$ &  $2.04\pm0.08$   &  --    & --  &  1936/1931   \\ 
\noalign{\smallskip}
\noalign{\smallskip}
F04454$-$4838 &  ESO\,203$-$IG001  &	--	& -- &  --   &  --    & --  &  --   \\ 
\noalign{\smallskip}
\noalign{\smallskip}
07251$-$0248	&    &	 -- 	& --  &   $ 4.4_{-1.8}^{+2.8}$  &  --    & --  &  9.1/12     \\ %
\noalign{\smallskip}
\noalign{\smallskip}
F08354+2555 & NGC\,2623 &	 $ \leq 1.4$ 	& $1.1^{+0.5}_{-0.4}$ &  $1.8^{\rm A}$   & $7.1^{+5.0}_{-2.8}$    & $1.8^{\rm B}$ &  571/569   \\ 
\noalign{\smallskip}
\noalign{\smallskip}
F08520$-$6850  & ESO\,060$-$IG16 (NE)   &	 $ 8.3_{-7.0}^{+6.5}$ 	& $0.12^{+1.09}_{-0.05}$ &  $1.8^{\rm A}$   & $15^{+10}_{-6}$ &   $1.8^{\rm B}$      &  457/512   \\ %
\noalign{\smallskip}
\noalign{\smallskip}
F08572+3915 &    &	 -- 	& --  &   $0.7\pm0.3$  &  --    & --  & 412/425    \\ %
\noalign{\smallskip}
\noalign{\smallskip}
F09111$-$1007	&   	 W	&   --  & -- 	  &  $1.4\pm0.6$   &  --    & --  &  32/25     \\ %
\noalign{\smallskip}
F09111$-$1007	&   	 E	&   --   	& --  &  $2.2\pm0.3$  &  --    & --  &  66/71     \\ %
\noalign{\smallskip}
\noalign{\smallskip}
F10038$-$3338 &   &	 $ 7_{-3}^{+8}$ 	& $0.17^{+0.16}_{-0.10}$ &  $1.55_{-0.85}^{+1.14}$   &  --    & --  &  74/71   \\ %
\noalign{\smallskip}
\noalign{\smallskip}
10565+2448 &   &	 $1.6_{-0.6}^{+0.7}$ 	& $0.78^{+0.10}_{-0.12}$ &  $1.98_{-0.26}^{+0.30}$   &  --    & --  & 536/617     \\ %
\noalign{\smallskip}
\noalign{\smallskip}
F12112+0305 &   &	 $\leq 0.9$ 	& $0.93^{+0.13}_{-0.11}$ &  $1.45_{-0.33}^{+0.45}$   &  --    & --  &   213/216   \\ %
\noalign{\smallskip}
\noalign{\smallskip}
F12243$-$0036  & NGC\,4418   &	 $ \leq 2.7$ 	& -- &  $1.65^{+0.81}_{-0.54}$   &  --    & --  &   55/67   \\ %
\noalign{\smallskip}
\noalign{\smallskip}
F13126+2453 & IC\,860   &	--	& -- &  $1.8^{\rm B}$   &  --    & --  &  22/17   \\ %
\noalign{\smallskip}
\noalign{\smallskip}
F14348$-$1447 &  NE \& SW   &	 $\leq 1.5$ 	& $0.94^{+0.36}_{-0.32}$ &  $1.42^{+0.41}_{-0.29}$   &  --    & --  &  375/401   \\ %
\noalign{\smallskip}
F14348$-$1447 &  NE   &	 -- 	& -- &  $1.8\pm0.7$   &  --    & --  &  26/21   \\ %
\noalign{\smallskip}
F14348$-$1447 &  SW   &	 -- 	& -- &  $1.2\pm0.5$   &  --    & --  &  32/27   \\ %
\noalign{\smallskip}
\noalign{\smallskip}
14378$-$3651 &   &	 $\leq 48$ 	& -- &  $2.7_{-1.7}^{+3.4}$   &  --    & --  & 36/59    \\ %
\noalign{\smallskip}
\noalign{\smallskip}
F15250+3608 &   &	 $\leq 1.2$ 	& $0.66^{+0.13}_{-0.37}$ &  $2.5_{-0.6}^{+1.2}$   &  --    & --  & 220/230    \\ %
\noalign{\smallskip}
\noalign{\smallskip}
F17207$-$0014 &   &	 $8.0^{+1.7}_{-1.9}  $ 	& $0.29^{+0.12}_{-0.07}$ &  $ 1.60_{-0.32}^{+0.31}$   &  --    & --  &  451/518     \\ %
\noalign{\smallskip}
\noalign{\smallskip}
F18293$-$3413  &   &	 $6.2_{-1.2}^{+0.9}$ 	& $0.73^{+0.06}_{-0.08}$ &  $2.0_{-0.21}^{+0.19}$   &  --    & --  &   1481/1570  \\ %
\noalign{\smallskip}
\noalign{\smallskip}
F19297$-$0406 &   &	 $ \leq 3.9$ 	& -- &  $2.5_{-0.6}^{+0.9}$   &  --    & --  & 71/60    \\ %
\noalign{\smallskip}
\noalign{\smallskip}
F20550+1655  & CGCG\,448$-$020E \& CGCG\,448$-$020W  &	 $ 0.4_{-0.2}^{+0.3}$ 	& $0.78^{+0.05}_{-0.06}$ &  $1.64_{-0.10}^{+0.11}$   &  --    & --  & 1058/1092    \\ %
\noalign{\smallskip}
  & CGCG\,448$-$020W   &	 $ \leq 5.5$ 	& $0.82^{+0.21}_{-0.22}$ &  --  &  --    & --  & 68/74    \\ %
\noalign{\smallskip}
  & CGCG\,448$-$020E   &	$0.28_{-0.27}^{+0.81}$ 	& $0.7^{+1.6}_{-0.3}$  &   $2.2^{+0.6}_{-0.8}$   &  --    & --  & 62/85  \\ %
\noalign{\smallskip}
\noalign{\smallskip}
F20551$-$4250   & ESO\,286$-$IG19    &	 $ \leq 0.7$ 	& $0.82^{+0.05}_{-0.08}$ &  $1.60_{-0.24}^{+0.29}$   &  --    & --  &  716/725   \\ 
\noalign{\smallskip}
\noalign{\smallskip}
F23128$-$5919   & ESO\,148$-$IG002   &	 $ \leq 0.1$ 	& $0.74^{+0.06}_{-0.08}$ &  $0.94_{-0.13}^{+0.12}$   &  --    & --  &  741/798   \\ %
\noalign{\smallskip}
\noalign{\smallskip}
F23365+3604   &    &	$\leq 3.7$  	& --  &  $1.33^{+0.96}_{-0.61}$   &  --  & --  &  14/26    \\ %
\noalign{\smallskip}
\hline
\noalign{\smallskip}
\multicolumn{8}{l}{{\bf Notes.} $^{\rm\,A}$: value of $\Gamma_{\rm\,bin.}$ fixed to that of the AGN continuum ($\Gamma$); $^{\rm\,B}$: photon index fixed.} \\
\end{tabular}
\end{center}
\end{table*}

\clearpage

\section{Individual sources}\label{sect:individualsources}
In the following we report details on the X-ray spectral fitting of all new observations analyzed here.\medskip

\noindent{\it $\star$ IRAS\,F00344$-$3349 (ESO\,350$-$IG038)}\newline
This late-stage merging galaxy is not detected by {\it NuSTAR}. The {\it Chandra} image shows an extended source, comprising three knots of star-formation, overlying both galaxies \citep{Torres-Alba:2018yj}. The X-ray emission from this object is soft, with no clear hard X-ray component, and the star-formation model can reproduce very well the overall X-ray spectrum. Since this is a very close merger ($d_{\rm sep}=1.1$\,kpc), we follow the strategy of \citet{Torres-Alba:2018yj}, and consider the X-ray emission for the whole system.\smallskip

\noindent{\it $\star$ IRAS\,F01053$-$1746 (IC\,1623A \& IC\,1623B)}\newline
The {\it Chandra} image of this advanced merger system shows an extended source in the 0.3--10\,keV band, which covers both galaxies. As reported by 
\citet{Garofali:2020eq}, the 0.3--30\,keV X-ray emission can be described by the superposition of several point sources and some diffuse emission, all ascribed to star-formation. The source is clearly detected by {\it NuSTAR}. In order to be consistent with the {\it XMM-Newton} and {\it NuSTAR} observations, we use an extraction radius of $20\arcsec$ for the {\it Chandra} observation, to encompass both sources. The overall X-ray spectrum is soft, and can be well reproduced by our star-formation model, consistent with \cite{Garofali:2020eq}.\smallskip

\noindent{\it $\star$ IRAS\,F04454$-$4838 (ESO\,203$-$IG001)}\newline
Neither of the two galaxies in this early merger are detected by {\it Chandra} or {\it NuSTAR}. ESO\,203$-$IG001 is the only object not detected by {\it Chandra} in \citet{Iwasawa:2011fk}.\smallskip

\noindent{\it $\star$ IRAS\,07251$-$0248 }\newline
{\it Chandra} shows a faint point source consistent with this advanced merger. The X-ray spectrum could be well fit by a simple power-law model.\smallskip
\noindent{\it $\star$ IRAS\,F08354+2555 (NGC\,2623)  }\newline
A hard point source is detected by {\it Chandra} coincident with the position of the advanced merger NGC\,2623 (see also \citealp{Torres-Alba:2018yj}). The source is also detected by both {\it XMM-Newton} and {\it NuSTAR}. As discussed in Ricci et al. (in prep.), this system shows clear [Ne\,V] emission from {\it Spitzer}/IRS spectra \citep{Inami:2013il}, which suggests that it hosts an AGN. Based on the {\it Chandra} hardness ratio, the source is also classified as a candidate obscured AGN by \citet{Torres-Alba:2018yj}. We therefore use our AGN model for the spectral fit, which was able to reproduce well the broad-band X-ray emission. We find that the AGN is only mildly obscured, and has one of the lowest column densities in our sample for an AGN in late-stage mergers ($N_{\rm H}=7.1^{+10.7}_{-3.1}\times 10^{22}\rm cm^{-2}$).\smallskip

\noindent{\it $\star$ IRAS\,F08520$-$6850  (ESO\,060$-$IG16 NE \& SW)}\newline
This advanced merger (stage $c$) is detected both by {\it NuSTAR} and {\it Chandra}. A compact point source is detected in the {\it Chandra} image, overlapping with the nucleus of the NE galaxy \citep{Iwasawa:2011fk}. The source shows [Ne\,V] emission in the MIR \citep{Inami:2013il}, and is classified as an AGN also considering the {\it Chandra} hardness ratio \citep{Iwasawa:2011fk}. The X-ray spectrum is well fit by the AGN model, with the X-ray source being obscured by a line-of-sight column density of $N_{\rm H}=1.5^{+1.0}_{-0.6}\times 10^{23}\rm\,cm^{-2}$, consistent with what was previously found by \citet{Iwasawa:2011fk} using {\it Chandra} data.\smallskip

\noindent{\it $\star$ IRAS\,F08572+3915}\newline
Only a faint detection of this double system is obtained by {\it Chandra} and {\it XMM-Newton}. {\it Chandra} shows a point-like hard X-ray component from the northwest nucleus \citep{Iwasawa:2011fk}. The source is not detected by {\it NuSTAR}, from which we could infer an upper limit on the 10--24\,keV luminosity of $\log (L_{\rm 10-24}/\rm erg\,s^{-1})\leq 41.13$. The combined {\it XMM-Newton}/{\it Chandra} spectra could be well fit by a simple power-law model, which returned a very low photon index ($\Gamma=0.7\pm0.3$). Extending this model to higher energies would result in a 10--24\,keV luminosity of $\log (L_{\rm 10-24}/\rm erg\,s^{-1})=41.46$, i.e. higher than the upper limit inferred from our {\it NuSTAR} observations, which suggests that this hard X-ray component is not associated to an obscured AGN.
\smallskip

\noindent{\it $\star$ IRAS\,F09111$-$1007	}\newline
A point source was detected consistent with each of the two galaxies of this early merger. In both cases the X-ray emission is rather faint, and it could be well fit by a simple power-law model.\smallskip
\smallskip

\noindent{\it $\star$ IRAS\,F10038$-$3338}\newline
A compact source is detected in the {\it Chandra} image, coincident with the position of this late-stage merger galaxy \citep{Iwasawa:2011fk}. The source is not detected in the {\it NuSTAR} observation, and the X-ray spectrum is well fit by the star-formation model.\smallskip

\noindent{\it $\star$ IRAS\,10565+2448}\newline
{\it Chandra} detects a point source coincident with the western member of this advanced merger \citep{Iwasawa:2011fk}. The source is not detected by {\it NuSTAR}, and the X-ray spectrum is accurately modelled using the star-formation model.\smallskip

\noindent{\it $\star$ IRAS\,F12112+0305}\newline
The {\it Chandra} image shows two sources, coincident with the two optical nuclei \citep{Iwasawa:2011fk}. The combined {\it Chandra}/{\it XMM-Newton} spectrum is well fit by our star-formation model. The source is not detected by {\it NuSTAR}.\smallskip

\noindent{\it $\star$ IRAS\,F12243$-$0036  (NGC\,4418)  }\newline
Two point sources are found at a distance of $\sim 1.5\arcsec$ from each other in the {\it Chandra} images, with the eastern source being brighter above $\sim 2$\,keV \citep{Torres-Alba:2018yj}. The source is not detected by {\it NuSTAR}, and the X-ray spectrum is well fit with our star-formation model.\smallskip

\noindent{\it $\star$ IRAS\,F13126+2453  (IC\,860) }\newline
The source is only faintly detected by {\it Chandra}, and is not detected by {\it NuSTAR}. Due to the low signal-to-noise ratio of the spectrum, we fit it using a simple power-law model, with the photon index fixed to $\Gamma=1.8$.\smallskip

\noindent{\it $\star$ IRAS\,F14348$-$1447 (NE \& SW)}\newline
The {\it Chandra} image shows some diffuse X-ray emission, together with two point sources, with the southern one being brighter \citep{Iwasawa:2011fk}. The source is not detected by {\it NuSTAR}, and our star-formation model can well represent the X-ray spectrum. We also looked at the individual properties of the two nuclei in the {\it Chandra} observations, selecting circular regions of $2\arcsec$ around the sources. Due to the low signal-to-noise ratio, the two spectra are fitted with a simple power-law model.
\smallskip

\noindent{\it $\star$ IRAS\,14378$-$3651}\newline
{\it Chandra} shows the presence of a point-like source consistent with the nucleus of the galaxy, plus some soft, extended, X-ray emission \citep{Iwasawa:2011fk}. The source is not detected by {\it NuSTAR}, and the X-ray emission is well represented by a power-law component.
\smallskip

\noindent{\it $\star$ IRAS\,F15250+3608}\newline
The source is detected by both {\it Chandra} and {\it XMM-Newton}, with the former showing a soft point source with a position consistent with that of the optical counterpart \citep{Iwasawa:2011fk}. IRAS\,F15250+3608 is not detected by {\it NuSTAR}, and its X-ray spectrum is well fit by the star-formation model.
\smallskip

\noindent{\it $\star$ IRAS\,F17207$-$0014}\newline
This $d$-stage, single nucleus merger, is detected by {\it Chandra}, exhibiting two peaks, with the southern one being harder and coinciding with the position of the optical nucleus of the system \citep{Iwasawa:2011fk}. The source is also detected by {\it XMM-Newton}, but it is not detected by {\it NuSTAR}. The star-formation model provides a good fit to the X-ray spectrum.
\smallskip

\noindent{\it $\star$ IRAS\,F18293$-$3413 }\newline
{\it Chandra} shows resolved X-ray emission both in the soft and hard band \citep{Iwasawa:2011fk} of this minor merger, classified as stage $N$ \citep{Ricci:2017aa}. The source is detected by both {\it XMM-Newton} and {\it NuSTAR}, albeit for the latter only in the 3--10\,keV band. The combined spectra can be well represented by the star-formation model plus an additional emission feature, associated with Fe\,XXV (e.g., \citealp{Iwasawa:2009vn}). The data show in fact an excess at $\sim 6.6$\,keV, which can be well represented by a Gaussian line with a width fixed to $\sigma=10$\,eV, and energy of $6.64^{+0.12}_{-0.11}$\,keV. The equivalent width of the line is $360^{+181}_{-162}$\,eV.  A detailed analysis of Fe\,K$\alpha$ emission lines in GOALS objects will be presented in a forthcoming dedicated paper (Iwasawa et al. in prep.).
\smallskip

\noindent{\it $\star$ IRAS\,F19297$-$0406}\newline
The soft X-ray emission detected by {\it Chandra} for this late merger with a single nucleus is extended, while the hard X-ray emission is compact \citep{Iwasawa:2011fk}. This object is not detected by {\it NuSTAR}, and we use a simple power-law model to reproduce its X-ray emission. 
\smallskip

\noindent{\it $\star$ IRAS\,F20550+1655 (CGCG\,448$-$020E \& CGCG\,448$-$020W)}\newline
This late merger, with two nuclei separated by $5\arcsec$, is detected by both {\it Chandra} and {\it XMM-Newton}, but is not detected by {\it NuSTAR}. {\it Chandra} shows some diffuse X-ray emission, with two rather compact hard X-ray sources \citep{Iwasawa:2011fk}. The X-ray spectrum is well fit by the star-formation model.
\smallskip

\noindent{\it $\star$ IRAS\,F20551$-$4250  (ESO\,286$-$IG19)  }\newline
This system is a late-stage merger with a single nucleus, which is clearly detected by {\it Chandra} and {\it XMM-Newton}, but it is not detected by {\it NuSTAR}. The soft X-ray emission observed in {\it Chandra} is elongated, consistent with a star-formation related origin, while the hard X-ray emission is point-like, with some fainter elongation \citep{Iwasawa:2011fk}. Our fit with the star-formation model leaves clear residuals around $\sim 6$\,keV. We tested our AGN model, fixing $\Gamma=1.8$ (e.g., \citealp{Ricci:2017if}), and found that it provides a significant improvement on the fit. However, the expected 10--24\,keV luminosity from this model ($L_{10-24}=1.3\times10^{42}\rm\,erg\,s^{-1}$) is above the upper limit obtained by our {\it NuSTAR} observations ($L_{10-24}\leq 3.2\times10^{41}\rm\,erg\,s^{-1}$). This suggests that the excess is not associated with an AGN. Including a Gaussian line to the star-formation model, with width fixed to 10\,eV, improved the fit; the energy of the line is $6.60^{+0.05}_{-0.06}$\,keV, which suggests emission from Fe\,XXV.\smallskip

\noindent{\it $\star$ IRAS\,F23128$-$5919  (ESO\,148$-$IG002)}\newline
This system is a late-stage merger, which is well detected by {\it Chandra} and {\it XMM-Newton}, and is not detected by {\it NuSTAR}. The {\it Chandra} image shows extended X-ray emission, which covers both galactic nuclei. The X-ray spectrum is well fit by our star-formation model. \cite{Franceschini:2003uq} report the presence of an AGN absorbed by a column density of $N_{\rm H}\sim 7\times10^{22}\rm\,cm^{-2}$, and \cite{Iwasawa:2011fk} discuss that the AGN might be associated with the Southern nucleus. We apply our AGN model, fixing $\Gamma=1.8$ (e.g., \citealp{Ricci:2017if}), finding a very similar column density ($N_{\rm H}=7^{+5}_{-2}\times10^{22}\rm\,cm^{-2}$). However, the expected observed 10--24\,keV luminosity from this model ($8.3\times10^{41}\rm\,erg\,s^{-1}$) is higher than the upper limit inferred by our {\it NuSTAR} observations ($L_{10-24}\leq 1.7\times10^{41}\rm\,erg\,s^{-1}$). This implies that, if an AGN is present in this system, it is significantly more obscured than what is reported by \citet{Franceschini:2003uq}, and the observed X-ray emission of this object is thus dominated by star-formation. We therefore used the star-formation model for this object.

\smallskip

\noindent{\it $\star$ IRAS\,F23365+3604}\newline
{\it Chandra} shows a point-like source coinciding with the nucleus of this late merger. The source is not detected by {\it NuSTAR}, and its X-ray spectrum is well fit by a simple power-law model.

\section*{Acknowledgements}
 
We thank the referee for their useful suggestions, which helped us improving the quality of the manuscript. We thank Chin-Shin Chang for useful comments on the manuscript. LCH was supported by the National Key R\&D Program of China (2016YFA0400702) and the  National Science Foundation of China (11721303, 11991052). CR acknowledges support from the Fondecyt Iniciacion grant 11190831. ET acknowledge support from CATA-Basal AFB-170002, FONDECYT Regular grant 1190818, ANID Anillo ACT172033 and Millennium Nucleus NCN19\_058 (TITANs). SA gratefully acknowledges funding from the European Research Council (ERC) under the European UnionÕs Horizon 2020 research and innovation programme (grant agreement No 789410). FEB acknowledges support from ANID - Millennium Science Initiative Program - ICN12\_009, CATA-Basal - AFB-170002, and FONDECYT Regular - 1190818 and 1200495. VU acknowledges funding support from NASA Astrophysics Data Analysis Program (ADAP) Grant 80NSSC20K0450. AMM acknowledges support from the National Science Foundation under Grant No. 2009416. KI acknowledges support by the Spanish MICINN under grant Proyecto/AEI/10.13039/501100011033 and ``Unit of excellence Mar\'ia de Maeztu 2020-2023'' awarded to ICCUB (CEX2019-000918-M). PA acknowledges financial support from ANID Millennium Nucleus NCN19-058 (TITANS) and the Max Planck Society through a Partner Group. HI acknowledges support from JSPS KAKENHI Grant Number JP19K23462. This work made use of data from the {\it NuSTAR} mission, a project led by the California Institute of Technology, managed by the Jet Propulsion Laboratory, and funded by the National Aeronautics and Space Administration. We thank the {\it NuSTAR} Operations, Software and Calibration teams for support with the execution and analysis of these observations.  This research has made use of the {\it NuSTAR} Data Analysis Software (\textsc{NuSTARDAS}) jointly developed by the ASI Science Data Center (ASDC, Italy) and the California Institute of Technology (Caltech, USA), and of the NASA/ IPAC Infrared Science Archive and NASA/IPAC Extragalactic Database (NED), which are operated by the Jet Propulsion Laboratory, California Institute of Technology, under contract with the National Aeronautics and Space Administration.

\section*{Data Availability}

The datasets generated and/or analysed in this study are available from the corresponding author
on reasonable request.
 
\bibliographystyle{mnras}
 \bibliography{ULIRGs_nustar_ref}
 \section*{Affiliations}

$^{1}$N\'ucleo de Astronom\'ia de la Facultad de Ingenier\'ia, Universidad Diego Portales, Av. Ej\'ercito Libertador 441, Santiago 22, Chile\\
$^{2}$Kavli Institute for Astronomy and Astrophysics, Peking University, Beijing 100871, People's Republic of China\\
$^{3}$National Radio Astronomy Observatory, 520 Edgemont Rd, Charlottesville, VA 22903, USA\\
$^{4}$George Mason University, Department of Physics \& Astronomy, MS 3F3, 4400 University Drive, Fairfax, VA 22030, USA\\
$^{5}$IPAC, California Institute of Technology, 1200 E. California Blvd., Pasadena, CA 91125, USA\\
$^{6}$Institut de Ci\`encies del Cosmos, Universitat de Barcelona, IEEC-UB, Mart\'i i Franqu\`es, 1, 08028 Barcelona, Spain\\
$^{7}$ICREA, Pg. Llu\'is Companys, 23, 08010 Barcelona, Spain\\
$^{8}$Clemson University, Kinard Laboratory of Physics, Clemson, SC, USA\\
$^{9}$Instituto de Astrof\'isica and Centro de Astroingenier\'ia, Facultad de F\'isica, Pontificia Universidad Cat\'olica de Chile, Casilla 306, Santiago 22, Chile\\
$^{10}$Millennium Institute of Astrophysics, Nuncio Monse–or S—tero Sanz 100, Providencia, Santiago, Chile\\
$^{11}$Space Science Institute, 4750 Walnut Street, Suite 205, Boulder, Colorado 80301, USA\\
$^{12}$Department of Astronomy, School of Physics, Peking University, Beijing 100871, China \\
$^{13}$Department of Space, Earth and Environment, Chalmers University of Technology, Onsala Space Observatory, SE-439 92, Onsala, Sweden\\
$^{14}$Instituto de F\'isica y Astronom\'ia, Facultad de Ciencias, Universidad de Valpara\'iso, Gran Bretana N1111, Playa Ancha, Valpara\'iso, Chile\\
$^{15}$Department of Physics, University of Crete, GR-71003, Heraklion, Greece\\
$^{16}$Institute of Astrophysics, Foundation for Research and Technology$-$Hellas, Heraklion, GR-70013, Greece\\
$^{17}$Department of Astronomy, University of Virginia, Charlottesville, VA 22904, USA\\
$^{18}$Department of Astronomy, Beijing Normal University, 100875 Beijing, China\\
$^{19}$Hiroshima Astrophysical Science Center, Hiroshima University, 1-3-1 Kagamiyama, Higashi-Hiroshima, Hiroshima 739-8526, Japan\\
$^{20}$Eureka Scientific, 2452 Delmer Street Suite 100, Oakland, CA 94602-3017, USA\\
$^{21}$European Southern Observatory, Karl-Schwarzschild-Strasse 2, D-85748 Garching, Germany\\
$^{22}$Department of Astronomy, University of Massachusetts at Amherst, Amherst, MA 01003, USA\\
$^{23}$Ritter Astrophysical Research Center University of Toledo Toledo, OH 43606, USA \\
$^{24}$ARC Centre of Excellence for All Sky Astrophysics in 3 Dimensions (ASTRO 3D) \\
$^{25}$Institute for Astronomy, 2680 Woodlawn Drive, University of Hawaii, Honolulu, HI 96822, USA\\
$^{26}$Jet Propulsion Laboratory, California Institute of Technology, 4800 Oak Grove Drive, MS 169-224, Pasadena, CA 91109, USA\\
$^{27}$Department of Physics and Astronomy, 4129 Frederick Reines Hall, University of California, Irvine, CA 92697, USA\\
$^{28}$Department of Astronomy, Kyoto University, Kitashirakawa-Oiwake-cho, Sakyo-ku, Kyoto 606-8502, Japan\\

 \end{document}